\documentclass[english,superscriptaddress,twocolumn]{revtex4-2}
\usepackage{graphicx} 
\usepackage{amsmath}
\usepackage{comment}
\usepackage{xcolor}
\usepackage{amssymb}
\usepackage[toc,page]{appendix}

\begin{document}

\title{Axial Symmetry Breaking and Chiral Separation Effect in Nambu-Jona-Lasinio Model}

\author{Xin-Nan Zhu}
\affiliation{Institute of Particle Physics and Key Laboratory of Quark and Lepton Physics (MOS), \\
Central China Normal University, Wuhan 430079, China}

\author{Xin-Li Sheng}
\email[Corresponding author: ]{sheng@fi.infn.it}
\affiliation{Shanghai Research Center for Theoretical Nuclear Physics, 
NSFC and Fudan University, Shanghai 200438, China}

\author{Shu-Yun Yang}
\affiliation{School of Artificial Intelligence in Medicine, Guilin Medical University, Guilin, 541199, China}

\author{Shijun Mao}
\affiliation{School of Physics, Xi'an Jiaotong University, Xi'an, Shaanxi 710049, China}

\author{Defu Hou}
\email[Corresponding author: ]{houdf@ccnu.edu.cn}
\affiliation{Institute of Particle Physics and Key Laboratory of Quark and Lepton Physics (MOS), \\
Central China Normal University, Wuhan 430079, China}

\begin{abstract}
We investigate the chiral separation effect (CSE) and the axial-partner
susceptibility splittings within the three-flavor Nambu--Jona-Lasinio (NJL)
model, providing two complementary aspects of axial dynamics in hot and dense
magnetized matter. We compare a field-dependent scalar coupling, which leads
to inverse magnetic catalysis (IMC), with a constant coupling, which exhibits
magnetic catalysis (MC). In the weak-field limit, the CSE conductivity is
suppressed at low temperature and approaches its chiral-limit value as the
temperature increases. Its magnetic-field dependence is reversed between the
IMC and MC scenarios. On the other hand, the susceptibility splittings
$\Delta_A = \chi_{\pi^0} - \chi_{\delta^0}$ and
$\Delta\chi = \chi_{\pi^0} - \chi_{\sigma_l}$ are suppressed at high
temperature or density, reflecting the substantial restoration of axial
symmetries. Their difference develops a pronounced peak near the cross-over
or first-order transition, implying that the $U(1)_A^{(0)}$ and
$U(1)_A^{(3)}$ symmetries are restored at different temperatures or chemical
potentials. Both the weak-field CSE and the IMC susceptibility splittings at
vanishing chemical potential agree qualitatively with lattice-QCD
calculations. We also provide model predictions for the susceptibility
splittings at finite density.
\end{abstract}

\maketitle

\section{Introduction}

Relativistic heavy-ion collisions provide a unique opportunity to investigate strongly-interacting matter under extreme conditions~\cite{Shuryak:1978ij,Bjorken:1982qr,Gyulassy:2004zy,STAR:2005gfr,Shuryak:2014zxa}. In non-central collisions, extremely strong but short-lived magnetic fields are generated in the central region of the quark-gluon plasma (QGP), whose magnitude can reach several $m_\pi^2$ ($\sim 10^{18}$ Gauss) in Au+Au collisions at the Relativistic Heavy Ion Collider (RHIC) and can be one order of magnitude larger in Pb+Pb collisions at the Large Hadron Collider (LHC)~\cite{Kharzeev:2007jp,Skokov:2009qp,Voronyuk:2011jd,Deng:2012pc}. The interplay between the QGP and such strong magnetic fields leads to a group of anomalous phenomena. Among them, the chiral magnetic effect (CME)~\cite{Kharzeev:2007tn,Kharzeev:2007jp, Fukushima:2008xe, Son:2012wh,Kharzeev:2013ffa, Kharzeev:2024zzm} and chiral separation effect (CSE)~\cite{ Son:2004tq, Metlitski:2005pr, Newman:2005as} represent two closely related transport phenomena that couple the vector and axial-vector currents in the medium. The CME is a vector current along the magnetic field direction induced by a nonzero axial chemical potential, whereas the CSE is an axial current at finite vector chemical potential. The coupling between them gives rise to a collective excitation propagating in the direction of the magnetic field, i.e., the chiral magnetic wave (CMW)~\cite{Kharzeev:2010gd, Burnier:2011bf}. Reviews of anomalous effects in strongly interacting matter can be found in Refs. \cite{ Huang:2015oca, Kharzeev:2015znc, Miransky:2015ava,Bzdak:2019pkr,Shen:2025unr}.

Over the past decades, extensive experimental searches for the CME signatures have been carried out \cite{STAR:2009wot,STAR:2009tro,ALICE:2012nhw,Zhao:2019hta,STAR:2021mii}, but no conclusive evidence has been established due to substantial background contributions. The measurement of the CSE is even more challenging because of the difficulty in distinguishing the chirality of final-state hadrons. The CMW, on the other hand, provides a possible way for confirming the CSE indirectly \citep{Kharzeev:2013ffa,Landsteiner:2016led}. On the theoretical side, recent lattice-QCD calculations demonstrate that the CSE conductivity is strongly suppressed in the hadronic phase and approaches its chiral-limit value as the temperature increases, suggesting that the CSE provides useful information on the chiral properties of the strongly interacting matter near the crossover region \cite{Velasco:2022gaw,Brandt:2023wgf}.

The axial current induced by the CSE is the Noether current associated with an axial transformation and therefore its conservation is closely related to the axial-symmetry breaking or restoration. For QCD with three massless quark flavors, the global symmetry is $SU(3)_V\otimes SU(3)_A\otimes U(1)_V$, while the flavor-singlet $U(1)_A$ symmetry can be broken by a nontrivial gluonic topology at quantum level \cite{Adler:1969gk,Bell:1969ts,tHooft:1976rip,Ioffe:2006ww}. The non-singlet axial symmetry, $SU(3)_A$, on the other hand, is spontaneously broken by the chiral condensate at low temperature, and is restored at sufficiently high temperatures. A background magnetic field further modifies the flavor structure of axial transformations, leading to an explicit breaking of the axial transformations mixing quarks with different electric charges. In particular, neutral subgroups $U(1)_{A}^{(3)}$ and $U(1)_{A}^{(8)}$, generated by $\lambda_3\gamma^5$ and $\lambda_8\gamma^5$, remain unbroken by the magnetic field in the chiral limit. Here $\lambda_3$ and $\lambda_8$ denote the third and eighth Gell-Mann matrices, which commute with the charge matrix. As will be shown later, the axial current relevant to the CSE is a linear combination of Noether currents corresponding to the $U(1)_{A}^{(3)}$ and $U(1)_{A}^{(8)}$ symmetries. This relation motivates a joint analysis of the CSE and the related axial symmetries. 

In the past few years, properties of chiral and axial symmetries at finite temperature have been investigated extensively in both effective models \cite{Costa:2004db,Ruivo:2011fg,Jiang:2015xqz,GomezNicola:2018pbx,Li:2019chs,Wang:2021dcy} and first-principles lattice-QCD calculations \cite{Aoki:2012yj,Buchoff:2013nra,Tomiya:2016jwr,Mazur:2018pjw,Ding:2020xlj,Ding:2021jtn}. The fate of these symmetries is usually characterized by in-medium meson mass splittings, the topological susceptibility, and susceptibility splittings between axial-partners. In particular, the splittings $\chi_{\pi^0}-\chi_{\delta^0}$ and $\chi_{\pi^0}-\chi_{\sigma_l}$ quantify the non-degeneracy of two neutral axial-partner channels. Existing NJL or lattice-QCD studies \cite{Wang:2021dcy,Ding:2026ewc} have found an inverse-magnetic-catalysis-like behavior for these splittings, but their properties at finite chemical potential remain unclear. The finite susceptibility splittings indicate that the corresponding axial symmetries are broken and therefore the axial currents are non-conserved according to the Noether theorem. The CSE, on the other hand, describes the equilibrium response of the axial current. Studying both in a common framework therefore provide complementary information on axial dynamics and thus important for systematically exploring the axial properties in heavy-ion collisions, where temperature, baryon density, and magnetic fields play important roles. 

The influence of an external magnetic field on the QCD phase structure has attached lots of interest in recent years. The lattice-QCD calculations of the physical pion mass have revealed the phenomenon of inverse magnetic catalysis (IMC) \cite{Bali:2011qj,Bali:2012zg,Bruckmann:2013oba,Bali:2014kia, Endrodi:2015oba,Endrodi:2019zrl,Ding:2022tqn,Hattori:2023egw, Endrodi:2024cqn}. 
At high temperature, the summed chiral condensate of light quarks is suppressed by the magnetic field and the pseudo-critical temperature $T_{pc}$ of chiral symmetry decreases with increasing magnetic field. Various mechanisms have been proposed, but the origin behind the IMC effect remains unclear \cite{Bruckmann:2013oba,Fukushima:2012kc,Mao:2016fha,Mao:2016lsr,Mao:2017tcf,Mao:2019avr,Mao:2022dqn,Chao:2013qpa,Farias:2014eca,Ferreira:2014kpa}.

This work is organized as follows. In Sec. \ref{sec:symmetry} we discuss the axial transformations and define the axial current as well as the susceptibility splittings. Quantitative calculations are based on the three-flavor NJL model, which is described in Sec. \ref{sec:NJLmodel}. Numerical results for the CSE and the susceptibility splittings are presented in Sec. \ref{sec:CSE} and Sec. \ref{sec:suscept}, respectively. We finally conclude our paper in Sec. \ref{sec:summary}. Throughout this paper, we set quark chemical potentials $\mu_u=\mu_d=\mu_s=\mu_q$, which are related to the baryon chemical potential through $\mu_\text{B}=3\mu_q$.

\section{Axial symmetry and axial current}\label{sec:symmetry}

We begin by specifying the axial transformation in the flavor space.
For the quark triplet $\psi=\left(u,d,s\right)^{T}$, we have
\begin{equation}
\psi\rightarrow e^{i\gamma^{5}\Theta}\psi\,,\ \ \ \ \overline{\psi}\rightarrow\overline{\psi}e^{i\gamma^{5}\Theta}\,,
\end{equation}
where the flavor matrix $\Theta$ can be generally decomposed
in terms of the $3\times3$ unit matrix $T_{0}=I_{3\times3}$ and
the Gell-Mann matrices $T_{a}=\lambda_{a}/2,\ a=1,\cdots,8$, 
\begin{equation}
\Theta=\alpha_{0}T_{0}+\sum_{a=1}^{8}\alpha_{a}T_{a}\,.
\end{equation}
The right-handed and left-handed components transform differently
as,
\begin{equation}
\psi_{R}\rightarrow e^{i\Theta}\psi_{R}\,,\ \ \ \ \psi_{L}\rightarrow e^{-i\Theta}\psi_{L}\,,
\end{equation}
where the off-diagonal elements of $\Theta$ mix quark flavors. The
generator $T_{0}$ defines the flavor-singlet $U(1)_{A}$ transformation,
whereas the traceless generators $T_{a}$ define flavor non-singlet
axial transformations belonging to $SU(3)_{A}$. The $T_{3}$ and
$T_{8}$ generators are particularly important in a magnetic field because they commute
with the quark charge matrix and thus does not mix flavors carrying different
electric charges. For the convenience of later discussion, we denote $U(1)_A^{(i)}$ 
as the transformation corresponding to the $i$-th generator $T_i$.

The Noether currents associated with the singlet and non-singlet generators
are defined as follows
\begin{equation}
J_{5}^{(0)\mu}=\left\langle\overline{\psi}\gamma^{\mu}\gamma^{5}\psi\right\rangle\,,\ \ \ \ J_{5}^{(a)\mu}=\left\langle\overline{\psi}\gamma^{\mu}\gamma^{5}\frac{\lambda_{a}}{2}\psi\right\rangle\,.
\end{equation}
For the NJL model introduced in the next section, the anomaly equations read
\begin{align}
\partial_{\mu}J_{5}^{(0)\mu}=&\   2i\left\langle\overline{\psi}m_{f}\gamma^{5}\psi\right\rangle+6iK\left\langle \text{det}_{f}\left[\overline{\psi}(1+\gamma_{5})\psi\right]\right. \nonumber\\
&\left.-\text{det}_{f}\left[\overline{\psi}\left(1-\gamma_{5}\right)\psi\right]\right\rangle\,, \nonumber \\
\partial_{\mu}J_{5}^{(a)\mu}  =&\   \frac{i}{2}\left\langle\overline{\psi}\left\{ m_{f},\lambda_{a}\right\} \gamma^{5}\psi\right\rangle\,,
\end{align}
where the term involving $K$ originates from the Kobayashi-Maskawa-'tHooft (KMT) interaction \cite{Kobayashi:1970ji, Kobayashi:1971qz, tHooft:1976rip, tHooft:1976snw}. Since gluons are not explicit degrees of freedom in the NJL model, the anomaly equations do not rely on the axial anomaly of gluon fields. Instead, the KMT interaction provides an effective description for such an anomaly and consequently the explicit $U(1)_{A}^{(0)}$ symmetry is broken even in the absence of magnetic field \citep{Fukushima:2001hr, Kunihiro:2009ds}. The KMT interaction does not contribute to the anomaly equation of $J_5^{(a)\mu}$ because it is invariant under the $SU(3)_{A}$ transformation. The quark mass, on the other hand, explicitly break both the singlet and the non-singlet axial symmetries.

We introduce the electric-charge-weighted axial current as 
\begin{equation}\label{eq:CSE-current}
J_{5\mu}^{Q}(x)\equiv\sum_{f=u,d,s}Q_{f}\left\langle \overline{\psi}_{f}(x)\gamma_{\mu}\gamma_{5}\psi_{f}(x)\right\rangle\,,
\end{equation}
where $Q_{u}=2/3$ and $Q_{d}=Q_{s}=-1/3$ are quark charges 
in units of $e$. We can further express $J_{5\mu}^{Q}(x)$ in terms
of Noether currents associated with flavor non-singlet $U(1)_{A}^{(3)}$ and $U(1)_{A}^{(8)}$
transformations, 
\begin{equation}
J_{5\mu}^{Q}(x)=J_{5\mu}^{(3)}(x)+\frac{1}{\sqrt{3}}J_{5\mu}^{(8)}(x)\,.
\end{equation}
The conservation of $J_{5\mu}^Q$ is therefore governed by the explicit breakings of $U(1)_{A}^{(3)}$ and $U(1)_{A}^{(8)}$ symmetries. In addition to the non-equilibrium evolution described
by the anomaly equations, $J_{5\mu}^Q$ contains the equilibrium response to the magnetic field in the presence of finite chemical potential $\mu_q$, i.e., the CSE. This response is linear in the magnetic field strength $|eB|$ and $\mu_q$ when both $|eB|$ and $\mu_q$ are small, allowing us to express it as follows,
\begin{equation}\label{eq:axial-current}
eJ_{5\mu}^{Q}(x)=\sigma_{\text{CSE}}(|eB|,\mu_q,T)\,\mu_q|eB|b_{\mu}\,,
\end{equation}
where the CSE conductivity $\sigma_{\text{CSE}}$ is in general a function of $|eB|$, $\mu_q$, and the temperature $T$. 

For the flavor singlet $U(1)_{A}^{(0)}$ and non-singlet $U(1)_{A}^{(8)}$ symmetries,
we find three sets of axial-partners, i.e., 
\begin{equation}
U(1)_{A}^{(0)} \text{ and } U(1)_{A}^{(8)}:\ \ \pi^{0}\leftrightarrow\delta^{0}\,,\ \ \eta_{l}\leftrightarrow\sigma_{l}\,,\ \ \eta_{s}\leftrightarrow\sigma_{s}\,,
\end{equation}
On the other hand, we have two sets of axial-partners for the flavor non-singlet
$U(1)_{A}^{(3)}$ symmetry,
\begin{equation}
U(1)_{A}^{(3)}:\ \ \ \pi^{0}\leftrightarrow\sigma_{l}\,,\ \ \ \ \eta_{l}\leftrightarrow\delta^{0}\,.
\end{equation}
The corresponding mesonic operators are defined as
\begin{eqnarray}
&&\pi^{0}=\frac{P_{u}-P_{d}}{\sqrt{2}}\,,\ \ \eta_{l}=\frac{P_{u}+P_{d}}{\sqrt{2}}\,,\ \  \eta_{s}=P_{s}\,,\nonumber \\
&&\delta^{0}=\frac{S_{u}-S_{d}}{\sqrt{2}}\,,\ \  \sigma_{l}=\frac{S_{u}+S_{d}}{\sqrt{2}}\,,\ \  \sigma_{s}=S_{s}\,,
\end{eqnarray}
where $S_f$ and $P_f$ are scalar and pseudoscalar quark bilinears 
\begin{equation}
S_{f}=\overline{\psi}_{f}\psi_{f},\ \ \ \ P_{f}\equiv i\overline{\psi}_{f}\gamma^{5}\psi_{f}\,.
\end{equation}
We note that $\eta_l$ and $\eta_s$ denote the light- and strange-flavor pseudoscalars rather than physical $\eta$ and $\eta^\prime$ mesons due to the mixing between $\eta$ and $\eta^\prime$. 
In order to quantify the axial symmetry breakings, we choose the following partner susceptibility splittings as probes,
\begin{align}\label{def:splittings}
\Delta_{A}=&\ \chi_{\pi^{0}}-\chi_{\delta^{0}}\,, \nonumber\\
\Delta_\chi=&\ \chi_{\pi^{0}}-\chi_{\sigma_l}\,,\nonumber\\
\Delta_\text{disc}=&\ \Delta_A-\Delta_\chi\,,
\end{align}
where the susceptibility of a mesonic operator $O$ is defined as
\begin{equation}\label{def:susceptibility}
\chi_{\mathcal{O}}\equiv\int d^{4}x\,\left\langle \mathcal{O}(x)\mathcal{O}(0)\right\rangle _{c}\,,
\end{equation}
with the connected mean value being
$\left\langle \mathcal{O}_{1}\mathcal{O}_{2}\right\rangle _{c}\equiv\left\langle \mathcal{O}_{1}\mathcal{O}_{2}\right\rangle -\left\langle \mathcal{O}_{1}\right\rangle \left\langle \mathcal{O}_{2}\right\rangle $.
The splittings $\Delta_{A}$, $\Delta_\chi$, and $\Delta_\text{disc}$ are associated with the surviving $U(1)_A^{(0)}$ (or $U(1)_A^{(8)}$)
and $U(1)_A^{(3)}$ symmetries and the difference between them, respectively. The last term arises from the flavor-disconnected contribution and thus reflects the correlation between $u$ and $d$ quark sectors. In the three-flavor case, the $\pi^0$-$\delta^0$ pair also transforms under the $T_8$ generator. As a consequence, one cannot distinguish $U(1)_A^{(0)}$ and $U(1)_A^{(8)}$ breakings using $\Delta_{A}$ alone. In this paper, we presume that the breaking and restoration of $U(1)_A^{(0)}$ and $U(1)_A^{(8)}$ are simultaneous and thus identify $\Delta_{A}$ as the probe of $U(1)_A^{(0)}$ symmetry breaking. The separation of $U(1)_A^{(8)}$ symmetry, which requires analyzing together with strange-quark sectors, are left for future studies.

\section{Three-flavor NJL model}\label{sec:NJLmodel}

For a strongly-interacting quark matter, we employ the three-flavor
NJL model with scalar four-fermion interactions and six-fermion KMT interaction \cite{Klevansky:1992qe,Hatsuda:1994pi,Buballa:2003qv,Volkov:2005kw},
\begin{align}
\mathcal{L}_{\text{eff}} =& \sum_{f=u,d,s}\overline{\psi}_{f}(i\gamma_{\mu}D_{f}^{\mu}-m_{f})\psi_{f}\nonumber\\
&+G_{S}\sum_{a=0}^{8}\left[(\overline{\psi}\lambda_{a}\psi)^{2}+(\overline{\psi}i\gamma_{5}\lambda_{a}\psi)^{2}\right]\nonumber \\
 & -K\left\{ \text{det}_{f}\left[\overline{\psi}(1+\gamma_{5})\psi\right]+\text{det}_{f}\left[\overline{\psi}\left(1-\gamma_{5}\right)\psi\right]\right\} \,,\label{eq:effective Lagrangian}
\end{align}
where $\psi=(\psi_{u},\psi_{d},\psi_{s})$ are Dirac spinors for $u$,
$d$, and $s$ quarks, $m_{f}$ denotes the current quark mass, $\lambda_{a}$
with $a=1,\cdots,8$ are Gell-Mann matrices in the flavor space, and
$\lambda_{0}=\sqrt{2/3}I_{3\times3}$. In an external
electromagnetic field, $D_{f}^{\mu}\equiv\partial^{\mu}+iQ_{f}eA^{\mu}$
is the covariant derivative with $Q_{f}$ the quark electric charge in units of $e$
and $A^{\mu}$ the gauge potential.

Under the mean-field approximation, we introduce the chiral condensates
for each flavor of quark
\begin{equation}
\sigma_{f}\equiv\left\langle \overline{\psi}_{f}\psi_{f}\right\rangle\,,\ \ \ \ f=u\,,d\,,s\,,\label{eq:Chiral-condensate}
\end{equation}
and the Lagrangian becomes 
\begin{eqnarray}
\mathcal{L}_{\text{MF}} & = & \sum_{f=u,d,s}\overline{\psi}_{f}\left(i\gamma_{\mu}D_{f}^{\mu}-M_{f}\right)\psi_{f}\nonumber \\
 &  & -2G_{S}\sum_{f=u,d,s}\sigma_{f}^{2}+4K\sigma_{u}\sigma_{d}\sigma_{s}\,,
\end{eqnarray}
where the dynamical quark mass $M_{f}$ is given by 
\begin{equation}
M_{f}\equiv m_{f}-4G_{S}\sigma_{f}+2K\prod_{f^{\prime}\neq f}\sigma_{f^{\prime}}\,.\label{eq:dynamic_mass}
\end{equation}
In a thermal equilibrium system, the grand thermodynamic potential
reads 
\begin{equation}
\Omega=\sum_{f=u,d,s}\left(2G_{S}\sigma_{f}^{2}-\Omega_{f}\right)-4K\sigma_{u}\sigma_{d}\sigma_{s}\,,\label{eq:thermodynamic_potential}
\end{equation}
where the quark contribution is given by 
\begin{align}
\Omega_{f}=&\frac{N_{c}|q_{f}B|}{4\pi^{2}}\sum_{n}d_{(n)}\int dp_{z}\bigg\{ E_{f}^{(n)}
\nonumber\\
&+T\sum_{s}\ln\left[1+\exp\left(-E_{f}^{(n)}/T+s\mu_{f}/T\right)\right]\bigg\}\,. \label{eq:quark_potential}
\end{align}
The degeneracy of color is $N_{c}=3$ and the degeneracy of Landau
level is $d_{(0)}=1$ and $d_{(n)}=2$ for $n\geq1$. For each flavor
of quark, the eigen-energies are determined by the Landau energy levels,
\begin{equation}
E_{f}^{(n)}=\sqrt{p_{z}^{2}+M_{f}^{2}+2n|Q_{f}eB|}\,,\ \ \ \ n\geq0\,.
\end{equation}
The chiral condensates (\ref{eq:Chiral-condensate}) are obtained
by minimizing $\Omega$ in Eq. (\ref{eq:thermodynamic_potential}),
\begin{equation}
\frac{\partial\Omega}{\partial\sigma_{u}}=\frac{\partial\Omega}{\partial\sigma_{d}}=\frac{\partial\Omega}{\partial\sigma_{s}}=0\,,
\end{equation}
which are also known as the gap equations for the NJL model.

Since the NJL model is non-renormalizable, it is necessary to introduce
a regularization scheme for the momentum integrals in Eq. (\ref{eq:quark_potential}).
Following Refs. \cite{Pauli:1949zm,Carignano:2019ivp}, we choose the Pauli-Villars regularization
scheme, under which a function of $M_{f}$ is replaced by 
\begin{equation}
f(M_{f})\rightarrow f_{\text{P.V.}}(M_{f})=\sum_{j=0}^{3}c_{j}f\left(\sqrt{M_{f}^{2}+j\Lambda^{2}}\right)\,,
\end{equation}
with $c_{0}=1$, $c_{1}=-3$, $c_{2}=3$, and $c_{3}=-1$. The parameters
in the model are chosen as follows in the absence of magnetic
field \cite{Carignano:2019ivp}, 
\begin{align}
&m_{u,d}=10.30\,\text{MeV},\quad m_{s}=236.9\,\text{MeV},\nonumber\\
&\Lambda=0.7812 \text{GeV},\quad G_{S}\Lambda^{2}=4.900,\quad K\Lambda^{5}=129.8, \label{eq:parameters}
\end{align}
which are obtained by fitting the vacuum values for the
pion decay constant and masses of pion, kaon, $\eta^{\prime}$, and
the constituent light-quark.

In a magnetic field, we consider two choices for the
scalar coupling $G_S$, the IMC scenario and the MC scenario. In the later case, $G_S$ is fixed to its vacuum value (\ref{eq:parameters}). This leads to enhancement of dynamical quark masses in magnetic fields as indicated by the MC effect. In the IMC scenario, we use a field-dependent value $G_S(|eB|)$ which is parameterized as
\begin{equation}
G_{S}(|eB|)=G_{S}(0)\frac{1+a\zeta^{2}+b\zeta^{3}}{1+c\zeta^{2}+d\zeta^{4}}\,,
\end{equation}
with $\zeta=|eB|/m_{\pi}^{2}$ and $m_{\pi}=0.138$ GeV. The parameters
$a=-5.645\times10^{-4}$, $b=7.107\times10^{-6}$, $c=8.340\times10^{-6}$,
and $d=5.944\times10^{-8}$ are determined by fitting $T_{c}(|eB|)/T_{c}(0)$
to the lattice-QCD results \cite{Bali:2011qj}. 
In the NJL model, the pseudo-critical temperature for the chiral phase transition is 
determined by solving
$\sum_{l=u,d}\partial^2\sigma_l(B,\mu_q,T)/\partial T^2=0$.
Such a running $G_{S}$ reproduces the IMC behavior for light quarks in magnetic fields while leaving the vacuum properties unchanged.

\section{Chiral separation effect}\label{sec:CSE}

In the presence of magnetic field, the Dirac fields are quantized
according to the Landau energy levels. A direct calculation shows
that the axial current is purely determined by the lowest Landau level, while
contributions from higher Landau levels vanish due to the degeneracy
of spin,
\begin{align}
eJ_{5\mu}^{\mathcal{Q}}=&\ b_{\mu}\sum_{f=u,d,s}\frac{N_{c}Q_{f}^{2}|eB|}{4\pi^{2}}\int dp_{z}\nonumber\\
&\times\left[n_{f}^{+}(E_{f}^{(0)})-n_{f}^{-}(E_{f}^{(0)})\right]\,,
\end{align}
where $E_{f}^{(0)}=\sqrt{p_{z}^{2}+M_{f}^{2}}$ is the energy for
the lowest Landau level and $n_{f}^{\pm}(E)=1/[1+e^{(E\mp\mu_q)/T}]$
is the Fermi-Dirac distribution. 
Following Eq. (\ref{eq:axial-current}), we introduce the CSE conductivity $\sigma_{\text{CSE}}(|eB|,\mu_q,T)$ as a function of the magnetic field strength $|eB|$, the chemical potential $\mu_q$, and the temperature $T$. In the limit of $|eB|\rightarrow0$ and $\mu_q\rightarrow0$, the conductivity reads
\begin{equation}\label{eq:Sigma-CSE-zero-field}
\sigma_{\text{CSE}}(0,0,T)={\displaystyle \sum_{f}}\frac{N_{c}Q_{f}^{2}}{4\pi^{2}T}\int dp_{z}(\cosh\beta E_{f}^{(0)}+1)^{-1}.
\end{equation}
For chiral fermions with zero masses, the CSE conductivity is $1/(2\pi^2)$, which is its chiral limit \citep{Metlitski:2005pr}. The conductivity in Eq. (\ref{eq:Sigma-CSE-zero-field}) approaches to a constant value in the limit of $M_{f}\rightarrow0$,
\begin{equation}\label{eq:chiral-limit}
\lim_{M_{f}\rightarrow0}\sigma_{\text{CSE}}(0,0,T)=\frac{1}{2\pi^{2}}\left(N_c\sum_f Q_f^2\right)\,,
\end{equation}
which is $1/(2\pi^2)$ multiplied with a factor arises from the sum over color and flavor. Without causing confusion, we still call the right-hand-side of Eq. (\ref{eq:chiral-limit}) as the chiral limit of $\sigma_\text{CSE}$ in the NJL model. Since the current quark masses in our setup are nonzero, this limit cannot be reached even at sufficiently high temperature or density.

\begin{figure}[tbh]
    \centering
    \includegraphics[width=0.45\textwidth]{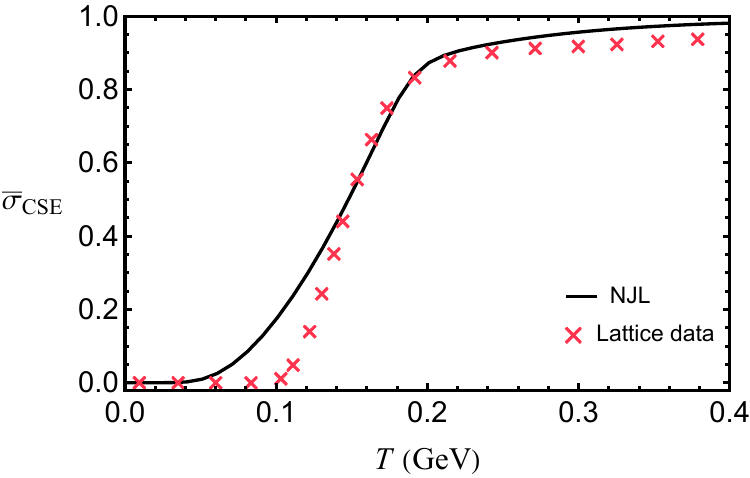}
    \caption{The normalized CSE conductivity $\bar\sigma_{\text{CSE}}$ as a function of temperature in the limit $|eB|\rightarrow0$ and $\mu_q\rightarrow0$. The solid line shows the result of the NJL model and the red crosses are lattice-QCD calculations~\cite{Brandt:2023wgf}.}
\label{fig:CSE}
\end{figure}

For the convenience of our discussion, we introduce the normalized conductivity
\begin{equation}
\bar{\sigma}_\text{CSE}\equiv\frac{2\pi^2}{N_c\sum_f Q_f^2}\sigma_\text{CSE}\,,
\end{equation}
whose chiral limit is normalized to 1. In Fig. \ref{fig:CSE}, we compare numerical results of $\bar{\sigma}_{\text{CSE}}$ within the three-flavor NJL model (the black solid line) with the lattice-QCD calculation \citep{Brandt:2023wgf}. The conductivity is strongly suppressed at low temperature and gradually increases toward the chiral limit at high temperatures. The NJL model agrees quantitatively with the lattice result at $T>0.16$ GeV. In particular, the conductivity is close to its chiral limit for $T>0.2$ GeV, which reflects the substantial restoration of the chiral symmetry at high temperatures. The NJL model fails to reproduce lattice-QCD results in the intermediate-temperature regime due to the lack of quark confinement. As shown in the Appendix, a mixed model composed by the NJL model and the hadron gas model can quantitatively reproduce the lattice results in the whole temperature range. 

\begin{figure}[tbh]
    \centering
\includegraphics[width=0.45\textwidth]{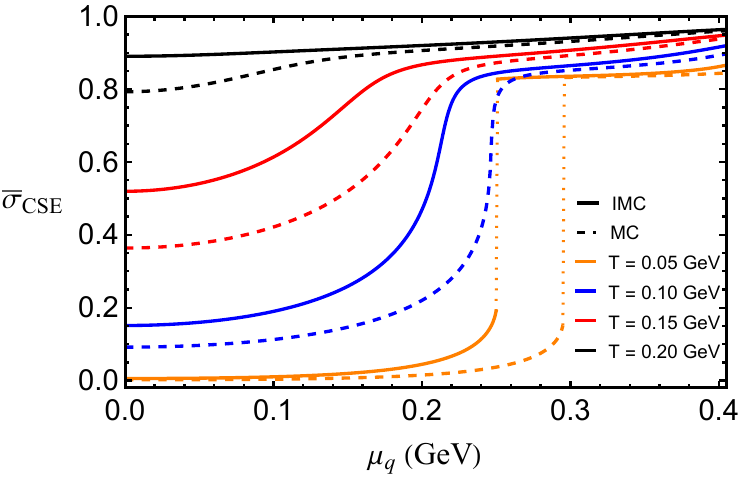}
    \caption{The normalized CSE conductivity $\bar{\sigma}_{\text{CSE}}$ as a function of the quark chemical potential $\mu_q$ at $eB=20\,m^{2}_{\pi}$ and $T=0.05$, 0.1, 0.15, and 0.2 GeV. Solid (dashed) lines denote results within the IMC (MC) scenario.}
\label{fig:sigma_mu}
\end{figure}

Figure \ref{fig:sigma_mu} shows $\bar{\sigma}_\text{CSE}$ as functions of the chemical potential at $|eB|=20\,m_\pi^2$ and $T=0.05$, 0.1, 0.15, and 0.2 GeV, respectively. At all considered temperatures, $\bar{\sigma}_\text{CSE}$ increases with $\mu_q$ approaching its chiral limit. At $T=0.05$ GeV with the IMC effect taken into account, the CSE is strongly suppressed at small $\mu_q$ due to large quark dynamic masses. It exhibits a discontinuous change at $\mu_q=0.25$ GeV towards its chiral limit, corresponding to a first-order chiral phase transition. At high temperatures, the curves are continuous and the slopes become gentle, as expected across a smooth crossover. The conductivity is enhanced by the temperature due to smaller quark dynamical masses at high temperatures. The conductivity in the MC scenario is smaller than that in the IMC scenario but the qualitative behaviors are similar. 
 
\begin{figure}[tbh]
    \centering
\includegraphics[width=0.45\textwidth]{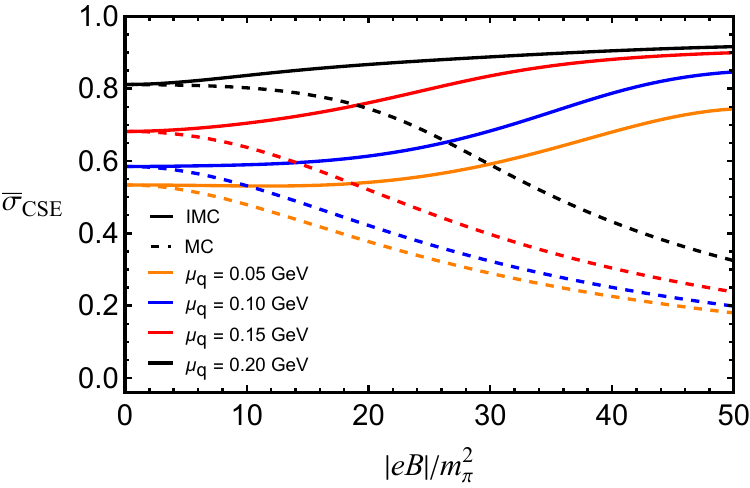}
    \caption{The normalized CSE conductivity $\bar{\sigma}_{\text{CSE}}$ as a function of the magnetic field strength at $T=0.15$ GeV. The orange, red, blue, and black curves correspond to $\mu_q = 0.05$, 0.1, 0.15, and 0.2 GeV, respectively. Solid (dashed) lines denote the results in the IMC (MC) scenario.}
\label{sigma_eB}
\end{figure}

The magnetic-field dependence of $\bar{\sigma}_\text{CSE}$ has also been studied and the results are shown in Fig. \ref{sigma_eB}. In the case of a field-dependence $G_S(|eB|)$, the IMC reduce quark masses as $|eB|$ increases and thus enhances $\bar{\sigma}_\text{CSE}$. In the MC scenario, the trend of $\bar{\sigma}_\text{CSE}$ is reversed.

\section{Susceptibility splittings of axial-partners}\label{sec:suscept}

The mesonic susceptibilities and the splittings in Eqs. (\ref{def:splittings}) and (\ref{def:susceptibility}) can be expressed as linear combinations of
$\left\langle S_{f_{1}}S_{f_{2}}\right\rangle _{c}$ and $\left\langle P_{f_{1}}P_{f_{2}}\right\rangle _{c}$. In the random phase approximation (RPA) \citep{Klevansky:1992qe,Hatsuda:1994pi,Buballa:2003qv}, the resummed correlations are evaluated by 
\begin{align}
\left\langle S_{f_{1}}S_{f_{2}}\right\rangle _{c}&=\left(\Pi_{S}+\Pi_{S}V_{S}\Pi_{S}+\cdots\right)_{f_{1}f_{2}} \nonumber\\
&=\left[\left(1-\Pi_{S}V_{S}\right)^{-1}\Pi_{S}\right]_{f_{1}f_{2}}\,,
\end{align}
and
\begin{align}
\left\langle P_{f_{1}}P_{f_{2}}\right\rangle _{c}&=\left(\Pi_{P}+\Pi_{P}V_{P}\Pi_{P}+\cdots\right)_{f_{1}f_{2}}\nonumber\\
&=\left[\left(1-\Pi_{P}V_{P}\right)^{-1}\Pi_{P}\right]_{f_{1}f_{2}}\,,
\end{align}
where $\Pi_{S}$ and $\Pi_{P}$ are one-loop polarization matrices
which are $3\times3$ and diagonal in the flavor space
\begin{align}\label{eq:one-loop-polarization}
\Pi_{ff}^{S} & \equiv -iN_{c} \int\frac{d^{4}q}{(2\pi)^{4}}\text{Tr}\left[G_{f}(q)G_{f}(q)\right]\,,\nonumber \\
\Pi_{ff}^{P} & \equiv  -iN_{c}\int\frac{d^{4}q}{(2\pi)^{4}}\text{Tr}\left[i\gamma_{5}G_{f}(q)i\gamma_{5}G_{f}(q)\right]\,,
\end{align}
and $V_{S}$, $V_{P}$ are RPA kernels for scalar and pseudoscalar channels,
\begin{align}
V_{S}&=\left(\begin{array}{ccc}
4G_{S} & -2K\sigma_{s} & -2K\sigma_{d}\\
-2K\sigma_{s} & 4G_{S} & -2K\sigma_{u}\\
-2K\sigma_{d} & -2K\sigma_{u} & 4G_{S}
\end{array}\right),\nonumber\\ 
V_{P}&=\left(\begin{array}{ccc}
4G_{S} & 2K\sigma_{s} & 2K\sigma_{d}\\
2K\sigma_{s} & 4G_{S} & 2K\sigma_{u}\\
2K\sigma_{d} & 2K\sigma_{u} & 4G_{S}
\end{array}\right)\,.
\end{align}
In Eq. (\ref{eq:one-loop-polarization}), $G_f(q)$ denotes the quark propagator in the background magnetic field. Detailed calculations of $\Pi_{ff}^S$ and $\Pi_{ff}^P$ are shown in Appendix \ref{sec:one-loop}. 
The susceptibility splittings $\Delta_A$, $\Delta_\chi$, and $\Delta_\text{disc}$ are then expressed as follows, 
\begin{align}
\Delta_A=&\ \left\langle P_u P_u+P_dP_d-2P_uP_d\right\rangle_c/2\nonumber\\
&-\left\langle S_u S_u+S_dS_d-2S_uS_d\right\rangle_c/2\,,\nonumber\\
\Delta_\chi=&\ \left\langle P_u P_u+P_dP_d-2P_uP_d\right\rangle_c/2\nonumber\\
&-\left\langle S_u S_u+S_dS_d+2S_uS_d\right\rangle_c/2\,, \nonumber\\
\Delta_\text{disc}=&\ 2\left\langle S_uS_d\right\rangle_c\,,
\end{align}
which will be used for later numerical calculations.

\begin{figure}[tb]
    \centering
\includegraphics[width=0.48\textwidth]{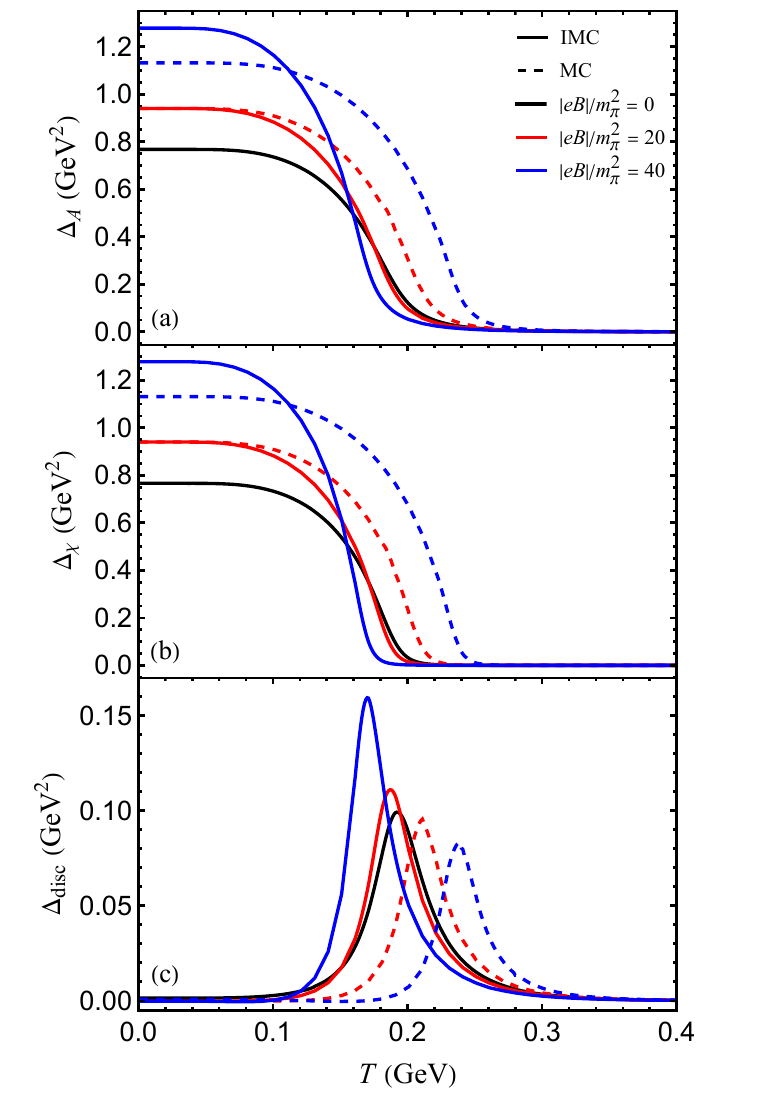}
    \caption{Susceptibility splittings $\Delta_A$, $\Delta_\chi$, and $\Delta_\text{disc}$ as functions of $T$ at $|eB|=0$, $20$, and $40\,m_\pi^2$ and $\mu_q=0$. Solid and dashed lines denote the IMC and MC scenarios, respectively. }
\label{fig:susceptibility-T}
\end{figure}

We present in Fig. \ref{fig:susceptibility-T} the temperature dependence of $\Delta_A$, $\Delta_\chi$, and $\Delta_\text{disc}$ at $\mu_q=0$ and $|eB|=0$, 20, and $40\,m_\pi^2$, respectively. In the IMC scenario, indicated by solid lines, both $\Delta_A$ and $\Delta_\chi$ increase with increasing $|eB|$ at low temperature. The ordering of three lines reverses at $T>0.18$ GeV, where a stronger magnetic field leads to smaller splittings. 
At an intermediate temperature, i.e., $0.16\,\text{GeV}<T<0.18\,\text{GeV}$, the magnetic-field dependence is non-monotonic: values of $\Delta_A$ and $\Delta_\chi$ at $|eB|=20\,m_\pi^2$ are larger than those at $|eB|=0$ or $40\,m_\pi^2$. In the MC scenario, by contrast, both $\Delta_A$ and $\Delta_\chi$ show clear enhancement by the magnetic field in the whole temperature range. 

The difference between $\Delta_A$ and $\Delta_\chi$, denoted as $\Delta_\text{disc}$, is shown in Fig. \ref{fig:susceptibility-T} (c) as a function of $T$. At low temperature, $\Delta_\text{disc}$ is nearly zero. It increases with $T$, reaches a peak near the crossover, and decreases to zero at high temperature. In the IMC (MC) scenario, the peak moves toward a lower (higher) temperature as $|eB|$ increases. 
The peak structure of $\Delta_\text{disc}$ indicates that restorations of $U(1)_A^{(0)}$ and $U(1)_A^{(3)}$ symmetries occur at different temperature regions. In the crossover region, the $U(1)_A^{(3)}$ symmetry already start to be restored, while the $U(1)_A^{(0)}$ symmetry breaking persists to higher temperatures.
 
\begin{figure}[tb]
    \centering
\includegraphics[width=0.48\textwidth]{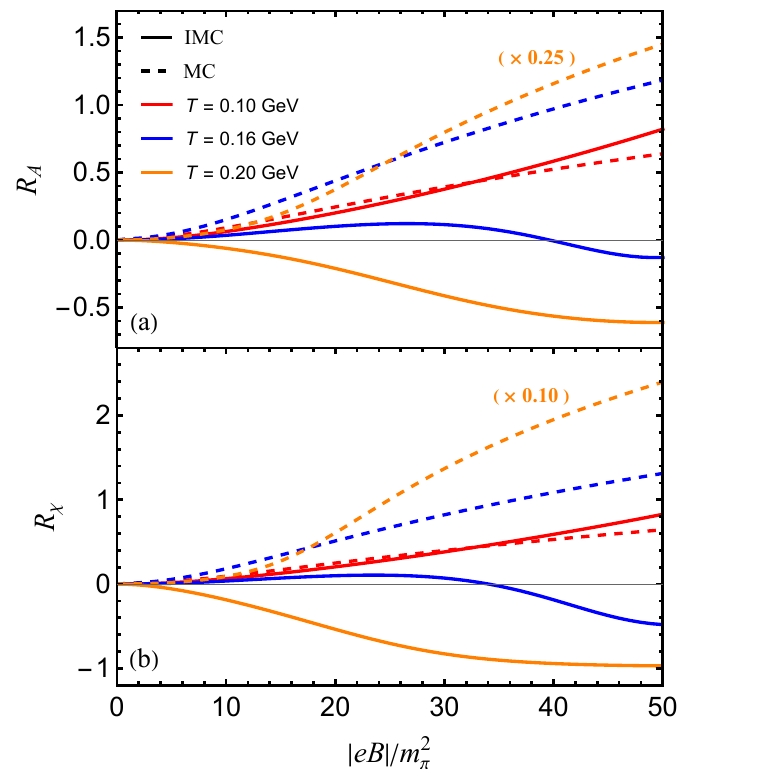}
\caption{Normalized magnetic responses $R_A$ and $R_\chi$ as functions of the magnetic field strength at $\mu_q=0$ and $T=0.1,\ 0.165$, and $0.2$ GeV, respectively. For visibility, the orange dashed lines are multiplied by 0.25 in panel (a) and by 0.1 in panel (b).}
\label{fig:suscept-2}
\end{figure}

\begin{figure}[tb]
    \centering
\includegraphics[width=0.48\textwidth]{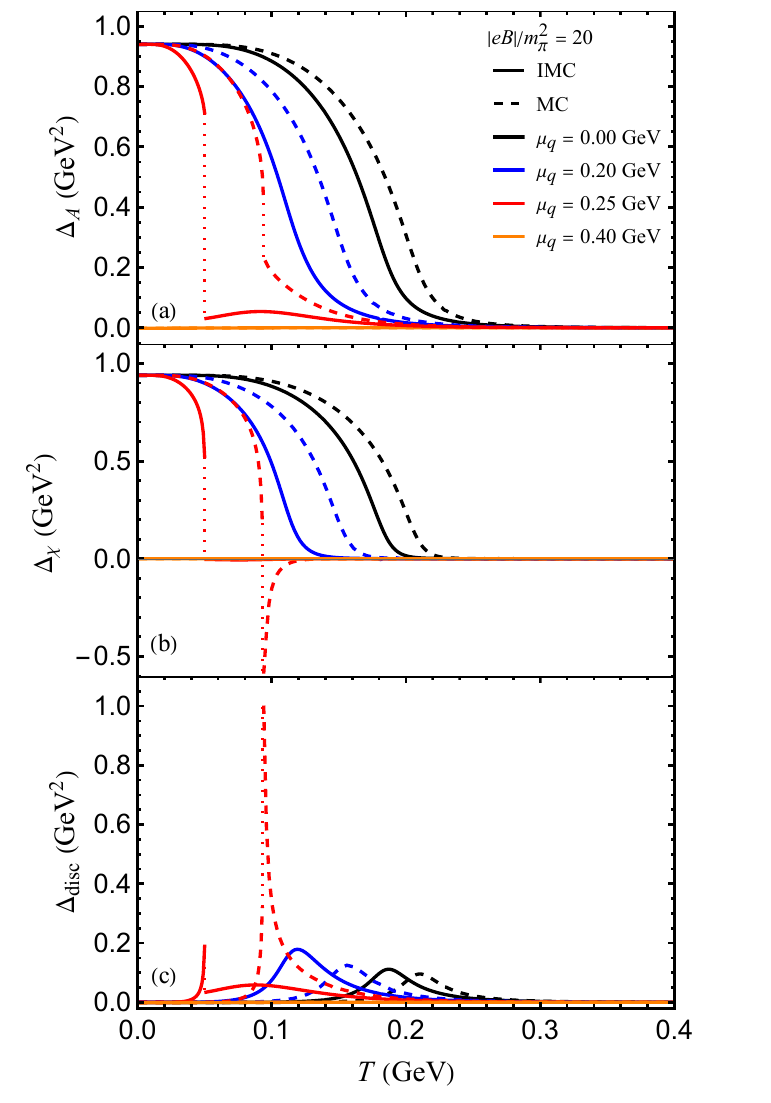}
    \caption{Susceptibility splittings $\Delta_A$, $\Delta_\chi$, and $\Delta_\text{disc}$ as functions of $T$ at $|eB|=20m_\pi^2$ and $\mu_q=0$, 0.2, 0.25, and 0.4 GeV. Solid (dashed) lines denote the IMC (MC) scenario.}
\label{fig:suscept-3}
\end{figure}

In order to quantify the magnetic-field dependence of $\Delta_A$ and $\Delta_\chi$, we show in Fig. \ref{fig:suscept-3} the normalized magnetic responses \citep{Ding:2026ewc}, 
\begin{equation}
R_{A}\equiv \frac{\Delta_A(|eB|,T)}{\Delta_A(0,T)}-1,\quad R_\chi\equiv\frac{\Delta_\chi(|eB|,T)}{\Delta_\chi(0,T)}-1\,.
\end{equation}
At $T=0.1$ GeV, these quantities are positive and increases at large $|eB|$ in both IMC and MC scenarios, as shown by the red solid and dashed lines. Things are very different at a high temperature, e.g., $T=0.2$ GeV. In the IMC scenario, the responses are negative and decreases with $|eB|$, while those in the MC scenario remain positive and  increases with $|eB|$. At an intermediate temperature, e.g., $0.16$ GeV, the curves in the IMC scenario are  nonmonotonic. As shown by the blue solid lines, they first increase and then decrease with $|eB|$. 

The suppression or enhancement of $\Delta_A$ and $\Delta_\chi$ at large magnetic field should be identified as susceptibility counterparts of the usual IMC or MC for light-quarks. When a field-dependent $G_S(|eB|)$ is applied to the NJL model, the crossover temperature is suppressed by the magnetic field and thus the $U(1)_A^{(0)}$ and  $U(1)_A^{(3)}$ symmetries start to be restored at lower temperatures. As a consequence, susceptibility splittings at fixed temperature will be smaller in a larger magnetic field. In the MC scenario when a constant $G_S$ is applied, on the other hand, the crossover temperature is raised by the magnetic field and consequently the susceptibility splittings are also enhanced. The qualitatively behavior of our NJL-model calculations in the IMC scenario agrees well with the recent lattice-QCD calculation in Ref. \cite{Ding:2026ewc}. 

We further extend our analysis to finite density. The temperature dependence of $\Delta_A$, $\Delta_\chi$, and $\Delta_\text{disc}$ at $\mu_q$=0, 0.2, 0.25, 0.4 GeV are shown in Fig. \ref{fig:suscept-3}, where the magnetic field strength is fixed to $|eB|=20\,m_\pi^2$. For $\mu_q=0$ and 0.2 GeV, both $\Delta_A$ and $\Delta_\chi$ gradually decrease with $T$ and become negligible at high temperature. When $\mu_q$ is sufficiently large, e.g., $\mu_q=0.4$ GeV as shown by the orange lines, both splittings are almost vanish in the whole temperature range. 
This indicates that both the $U(1)_A^{(0)}$ and the $U(1)_A^{(3)}$ symmetries can substantially restored at high chemical potential or high temperature.
At an intermediate $\mu_q$ such as $\mu_q=0.25$ GeV, we observe discontinuous drops of $\Delta_A$ and $\Delta_\chi$, corresponding to a first-order phase transition. At $\mu_q=0.25$ GeV in the MC scenario, $\Delta_\chi$ becomes negative near $T\sim0.1$ GeV, indicating that $\chi_{\sigma_l}>\chi_{\pi^0}$ in this region. Except this region, $\Delta_A$ and $\Delta_\chi$ in the IMC scenario are larger than those in the MC scenario, while their qualitative behaviors are similar.

The flavor-disconnected quantity, $\Delta_\text{disc}$, is shown in Fig. \ref{fig:suscept-3} (c) as the difference between $\Delta_A$ and $\Delta_\chi$ in panels (a) and (b). It is nearly zero at very low temperature, peaks near the crossover or the first-order phase transition, and vanishes at sufficiently high temperature. As $\mu_q$ increases, the peak shifts toward lower temperatures as a consequence of lower $T_c$ at higher $\mu_q$.

\begin{figure}[tb]
    \centering
\includegraphics[width=0.47\textwidth]{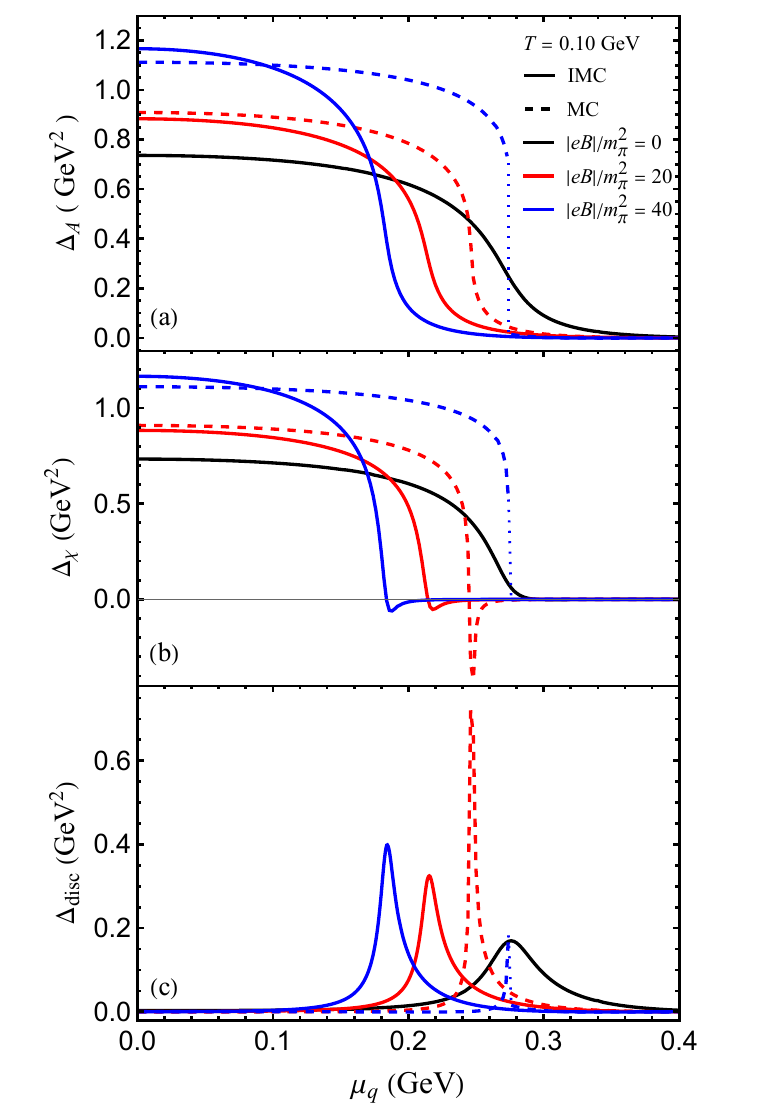}
    \caption{Susceptibility splittings $\Delta_A$, $\Delta_\chi$, and $\Delta_\text{disc}$ as functions of $\mu_q$ at $T=0.1$ GeV, in a magnetic field with strength $|eB|=0$, 20, and 40 $m_\pi^2$, respectively. Solid (dashed) lines denote the IMC (MC) scenario.}
\label{fig:suscept-4}
\end{figure}

Figure \ref{fig:suscept-4} presents the $\mu_q$-dependence of $\Delta_A$, $\Delta_\chi$, and $\Delta_\text{disc}$ at fixed temperature $T=0.1$ GeV and $|eB|=0$, 20, and 40 $m_\pi^2$. At low density, both $\Delta_A$ and $\Delta_\chi$ are enhanced by increasing magnetic field. Such an ordering reverses at high $\mu_q$ in the IMC scenario. In the MC scenario, on the other hand, $\Delta_A$ and $\Delta_\chi$ at $|eB|=40 m_\pi^2$ are always larger than those at $|eB|=20 m_\pi^2$. The disconnected term $\Delta_\text{disc}$ peaks near the cross-over or the first-order phase transition position. The peak position in the IMC scenario shifts toward lower $\mu_q$ at larger $|eB|$ and its height increases, while the field-dependence of the peak position exhibits a non-monotonic behavior in the other scenario. Together with the temperature scan, results of $\Delta_\text{disc}$ suggest that the $U(1)_A^{(0)}$ symmetry breaking persists to higher chemical potential or higher temperature than the $U(1)_A^{(3)}$ symmetry.

\section{Summary}\label{sec:summary}

In this work, we have investigated the CSE and the susceptibility splitting of axial-partners within the three-flavor NJL model in  hot and dense matter under external magnetic fields. These two effects are two aspects of the axial symmetry. The CSE describes the axial current as an equilibrium response to the magnetic field, while the susceptibility splittings quantify how the axial symmetries are restored. For the scalar coupling strength $G_S$ in the NJL model, we consider both the IMC and the MC scenarios. The former one uses a field-dependent $G_S(|eB|)$ fitted to reproduce the IMC while the second one uses a constant $G_S$ that produces the normal MC effect.

In the weak field and small $\mu_q$ limit, the CSE conductivity we obtain is nearly zero at low temperature. It increases with $T$ approaching its chiral limit. Above $T\simeq0.16$ GeV, the NJL model calculation quantitatively agrees with lattice-QCD calculations \citep{Brandt:2023wgf}. However, the NJL model overestimates the CSE at intermediate temperature due to the lack of quark confinement. Increasing $\mu_q$ will enhance the CSE by reducing the dynamical quark masses. In the IMC scenario, the CSE conductivity increases with $|eB|$, while it decreases in the MC scenario.

In order to investigate how the axial symmetries are broken or restored in the hot and dense magnetized matter, we focus on two susceptibility splittings, $\Delta_A=\chi_{\pi_0}-\chi_{\delta_0}$ and $\Delta_\chi=\chi_{\pi_0}-\chi_{\sigma_l}$, associated with  $U(1)_{A}^{(0)}$ and $U(1)_{A}^{(3)}$ symmetries, respectively. Both $\Delta_A$ and $\Delta{\chi}$ become small at high temperature or density, while their difference peaks around the crossover or the first-order chiral phase transition. This demonstrates that the restorations of the $U(1)_{A}^{0}$ and $U(1)_{A}^{3}$ symmetries occur at different temperature region. Since $\Delta_\text{disc}$ is positive, $\Delta_\chi$ is generally suppressed before $\Delta_A$, implying that the $U(1)_A^{(0)}$ persists to higher chemical potential or higher temperature.

At $\mu_q=0$, the magnetic field enhances $\Delta_A$ and $\Delta{\chi}$ at low temperatures. 
A field-dependent $G_S$ in the NJL model leads to the IMC effect for light quarks as well as suppression of these splittings at large magnetic field, which is identified as susceptibility counterparts of the IMC effect. Such behavior qualitatively agrees with the lattice-QCD calculations \citep{Ding:2026ewc}. In the MC scenario, both $\Delta_A$ and $\Delta{\chi}$ increase with $|eB|$. 
We further extend our work to finite density. Both $\Delta_A$ and $\Delta{\chi}$ become nearly zero at large $\mu_q$, indicating substantial restorations of both $U(1)_A^{(0)}$ and $U(1)_A^{(3)}$ symmetries. The flavor-disconnected quantity, $\Delta_\text{disc}$, still peaks around the cross-over or the first-order transition. The peak position shifts toward lower temperature as $\mu_q$ increases. The patterns of susceptibility splittings at fixed $T$ and varying $\mu_q$ are similar to those at $\mu_q=0$ and varying $T$. In the IMC scenario, the magnetic field will enhance $\Delta_A$ and $\Delta_\chi$ at small $\mu_q$ while suppress them at large $\mu_q$. In the MC scenario, in contrast, these splittings are enhanced by the magnetic field in the whole $\mu_q$ range.

The finite-density results are model predictions and show that the CSE and axial-partner susceptibility splittings provide complementary probes of the axial dynamics. These studies are particularly important for heavy-ion collisions at low energies, in which the QGP has a finite baryon chemical potential. A future extension to include the Polyakov loop \cite{Fukushima:2003fw, Ratti:2005jh, Roessner:2006xn}  would help effectively describe the quark confinement and thus improve the investigations at intermediate temperature.

\textit{Note added.—} When finalizing this work, we became aware of \citep{sheng2026exp}, which appeared very recently and presents an independent study of the $U_A(1)$ symmetry restoration within the NJL model. We note that the magnetic-field dependence of the susceptibility splittings found in \citep{sheng2026exp} is qualitatively similar to that obtained in the present work. 

\section*{ACKNOWLEDGMENTS}
This work is supported in part by the National Key Research and Development Program of China under Contract No. 2022YFA1604900.  D. H. is also supported by the National Natural Science Foundation of China(NSFC) under Grants No. 12435009, and No. 12275104. X.L.S. is supported by the National Natural Science Foundation of China under Grant No. 12547102 and No. 12147101. S.Y.Y. is supported by the Guangxi Young and Middle-aged University Teachers' (Research) Basic Competency Improvement Project (No. 2025KY0533). S.M. is supported by the National Natural Science Foundation of China under Grant No.12275204 and Natural Science Basic Research Plan in Shaanxi Province of China (Program No. 2026JC-YXQN-004). The data that support the findings of this article are openly available \cite{Zhu2026Dataset} .

\bibliographystyle{apsrev}
\bibliography{CSE-coefficient}

@misc{sheng2026exp,
      title={Exploring $U_A(1)$ symmetry restoration in a magnetic field by meson screening masses in the NJL model}, 
      author={Bing-Kai Sheng and Zigeng Ding and Danning Li and Xinyang Wang and Lang Yu},
      year={2026},
      eprint={2609.35468},
      archivePrefix={arXiv},
      primaryClass={hep-ph},
      url={https://arxiv.org/abs/2609.35468}, 
}

@article{Skokov:2009qp,
    author = "Skokov, V. and Illarionov, A. Yu. and Toneev, V.",
    title = "{Estimate of the magnetic field strength in heavy-ion collisions}",
    eprint = "0907.1396",
    archivePrefix = "arXiv",
    primaryClass = "nucl-th",
    doi = "10.1142/S0217751X09047570",
    journal = "Int. J. Mod. Phys. A",
    volume = "24",
    pages = "5925--5932",
    year = "2009"
}

@article{Deng:2012pc,
    author = "Deng, Wei-Tian and Huang, Xu-Guang",
    title = "{Event-by-event generation of electromagnetic fields in heavy-ion collisions}",
    eprint = "1201.5108",
    archivePrefix = "arXiv",
    primaryClass = "nucl-th",
    doi = "10.1103/PhysRevC.85.044907",
    journal = "Phys. Rev. C",
    volume = "85",
    pages = "044907",
    year = "2012"
}

@article{Shen:2025unr,
    author = "Shen, Diyu and Chen, Jinhui and Huang, Xu-Guang and Ma, Yu-Gang and Tang, Aihong and Wang, Gang",
    title = "{A Review of Intense Electromagnetic Fields in Heavy-Ion Collisions: Theoretical Predictions and Experimental Results}",
    eprint = "2512.00739",
    archivePrefix = "arXiv",
    primaryClass = "nucl-ex",
    doi = "10.34133/research.0726",
    journal = "Research",
    volume = "8",
    pages = "0726",
    year = "2025"
}

@article{Kharzeev:2007jp,
    author = "Kharzeev, Dmitri E. and McLerran, Larry D. and Warringa, Harmen J.",
    title = "{The Effects of topological charge change in heavy ion collisions: 'Event by event P and CP violation'}",
    eprint = "0711.0950",
    archivePrefix = "arXiv",
    primaryClass = "hep-ph",
    doi = "10.1016/j.nuclphysa.2008.02.298",
    journal = "Nucl. Phys. A",
    volume = "803",
    pages = "227--253",
    year = "2008"
}

@article{Fukushima:2008xe,
    author = "Fukushima, Kenji and Kharzeev, Dmitri E. and Warringa, Harmen J.",
    title = "{The Chiral Magnetic Effect}",
    eprint = "0808.3382",
    archivePrefix = "arXiv",
    primaryClass = "hep-ph",
    doi = "10.1103/PhysRevD.78.074033",
    journal = "Phys. Rev. D",
    volume = "78",
    pages = "074033",
    year = "2008"
}

@article{Kharzeev:2015znc,
    author = "Kharzeev, D. E. and Liao, J. and Voloshin, S. A. and Wang, G.",
    title = "{Chiral magnetic and vortical effects in high-energy nuclear collisions{\textemdash}A status report}",
    eprint = "1511.04050",
    archivePrefix = "arXiv",
    primaryClass = "hep-ph",
    doi = "10.1016/j.ppnp.2016.01.001",
    journal = "Prog. Part. Nucl. Phys.",
    volume = "88",
    pages = "1--28",
    year = "2016"
}

@article{Kharzeev:2024zzm,
    author = "Kharzeev, Dmitri E. and Liao, Jinfeng and Tribedy, Prithwish",
    title = "{Chiral magnetic effect in heavy ion collisions: The present and future}",
    eprint = "2405.05427",
    archivePrefix = "arXiv",
    primaryClass = "nucl-th",
    doi = "10.1142/9789811294679_0006",
    journal = "Int. J. Mod. Phys. E",
    volume = "33",
    number = "09",
    pages = "2430007",
    year = "2024"
}

@article{Newman:2005as,
    author = "Newman, G. M. and Son, D. T.",
    title = "{Response of strongly-interacting matter to magnetic field: Some exact results}",
    eprint = "hep-ph/0510049",
    archivePrefix = "arXiv",
    reportNumber = "INT-PUB-05-22",
    doi = "10.1103/PhysRevD.73.045006",
    journal = "Phys. Rev. D",
    volume = "73",
    pages = "045006",
    year = "2006"
}

@article{Metlitski:2005pr,
    author = "Metlitski, Max A. and Zhitnitsky, Ariel R.",
    title = "{Anomalous axion interactions and topological currents in dense matter}",
    eprint = "hep-ph/0505072",
    archivePrefix = "arXiv",
    doi = "10.1103/PhysRevD.72.045011",
    journal = "Phys. Rev. D",
    volume = "72",
    pages = "045011",
    year = "2005"
}

@article{Son:2004tq,
    author = "Son, D. T. and Zhitnitsky, Ariel R.",
    title = "{Quantum anomalies in dense matter}",
    eprint = "hep-ph/0405216",
    archivePrefix = "arXiv",
    reportNumber = "INT-PUB-04-13",
    doi = "10.1103/PhysRevD.70.074018",
    journal = "Phys. Rev. D",
    volume = "70",
    pages = "074018",
    year = "2004"
}

@article{Kharzeev:2013ffa,
    author = "Kharzeev, Dmitri E.",
    title = "{The Chiral Magnetic Effect and Anomaly-Induced Transport}",
    eprint = "1312.3348",
    archivePrefix = "arXiv",
    primaryClass = "hep-ph",
    doi = "10.1016/j.ppnp.2014.01.002",
    journal = "Prog. Part. Nucl. Phys.",
    volume = "75",
    pages = "133--151",
    year = "2014"
}

@article{Velasco:2022gaw,
    author = "Velasco, Eduardo Garnacho and Brandt, Bastian B. and Cuteri, Francesca and Endr{\H{o}}di, Gergely and Mark{\'o}, Gergely",
    title = "{Anomalous transport phenomena on the lattice}",
    eprint = "2212.02148",
    archivePrefix = "arXiv",
    primaryClass = "hep-lat",
    doi = "10.22323/1.430.0173",
    journal = "PoS",
    volume = "LATTICE2022",
    pages = "173",
    year = "2023"
}

@article{Brandt:2023wgf,
    author = "Brandt, Bastian B. and Endr{\H{o}}di, Gergely and Garnacho-Velasco, Eduardo and Mark{\'o}, Gergely",
    title = "{The chiral separation effect from lattice QCD at the physical point}",
    eprint = "2312.02945",
    archivePrefix = "arXiv",
    primaryClass = "hep-lat",
    doi = "10.1007/JHEP02(2024)142",
    journal = "JHEP",
    volume = "02",
    pages = "142",
    year = "2024"
}

@article{Klevansky:1992qe,
    author = "Klevansky, S. P.",
    title = "{The Nambu-Jona-Lasinio model of quantum chromodynamics}",
    doi = "10.1103/RevModPhys.64.649",
    journal = "Rev. Mod. Phys.",
    volume = "64",
    pages = "649--708",
    year = "1992"
}

@article{Buballa:2003qv,
    author = "Buballa, Michael",
    title = "{NJL model analysis of quark matter at large density}",
    eprint = "hep-ph/0402234",
    archivePrefix = "arXiv",
    doi = "10.1016/j.physrep.2004.11.004",
    journal = "Phys. Rept.",
    volume = "407",
    pages = "205--376",
    year = "2005"
}

@article{Volkov:2005kw,
    author = "Volkov, M. K. and Radzhabov, A. E.",
    title = "{The Nambu-Jona-Lasinio model and its development}",
    eprint = "hep-ph/0508263",
    archivePrefix = "arXiv",
    doi = "10.1070/PU2006v049n06ABEH005905",
    journal = "Phys. Usp.",
    volume = "49",
    pages = "551--561",
    year = "2006"
}

@article{Miransky:2015ava,
    author = "Miransky, Vladimir A. and Shovkovy, Igor A.",
    title = "{Quantum field theory in a magnetic field: From quantum chromodynamics to graphene and Dirac semimetals}",
    eprint = "1503.00732",
    archivePrefix = "arXiv",
    primaryClass = "hep-ph",
    doi = "10.1016/j.physrep.2015.02.003",
    journal = "Phys. Rept.",
    volume = "576",
    pages = "1--209",
    year = "2015"
}

@article{Hatsuda:1994pi,
    author = "Hatsuda, Tetsuo and Kunihiro, Teiji",
    title = "{QCD phenomenology based on a chiral effective Lagrangian}",
    eprint = "hep-ph/9401310",
    archivePrefix = "arXiv",
    reportNumber = "UTHEP-270, RYUTHP-94-1",
    doi = "10.1016/0370-1573(94)90022-1",
    journal = "Phys. Rept.",
    volume = "247",
    pages = "221--367",
    year = "1994"
}

@article{tHooft:1976rip,
    author = "'t Hooft, Gerard",
    editor = "Shifman, Mikhail A.",
    title = "{Symmetry Breaking Through Bell-Jackiw Anomalies}",
    reportNumber = "PRINT-76-0254 (HARVARD)",
    doi = "10.1103/PhysRevLett.37.8",
    journal = "Phys. Rev. Lett.",
    volume = "37",
    pages = "8--11",
    year = "1976"
}

@article{Pauli:1949zm,
    author = "Pauli, W. and Villars, F.",
    title = "{On the Invariant regularization in relativistic quantum theory}",
    doi = "10.1103/RevModPhys.21.434",
    journal = "Rev. Mod. Phys.",
    volume = "21",
    pages = "434--444",
    year = "1949"
}

@article{Carignano:2019ivp,
    author = "Carignano, Stefano and Buballa, Michael",
    title = "{Inhomogeneous chiral condensates in three-flavor quark matter}",
    eprint = "1910.03604",
    archivePrefix = "arXiv",
    primaryClass = "hep-ph",
    doi = "10.1103/PhysRevD.101.014026",
    journal = "Phys. Rev. D",
    volume = "101",
    number = "1",
    pages = "014026",
    year = "2020"
}

@article{Kharzeev:2007tn,
    author = "Kharzeev, D. and Zhitnitsky, A.",
    title = "{Charge separation induced by P-odd bubbles in QCD matter}",
    eprint = "0706.1026",
    archivePrefix = "arXiv",
    primaryClass = "hep-ph",
    reportNumber = "BNL-NT-07-24",
    doi = "10.1016/j.nuclphysa.2007.10.001",
    journal = "Nucl. Phys. A",
    volume = "797",
    pages = "67--79",
    year = "2007"
}

@article{Burnier:2011bf,
    author = "Burnier, Yannis and Kharzeev, Dmitri E. and Liao, Jinfeng and Yee, Ho-Ung",
    title = "{Chiral magnetic wave at finite baryon density and the electric quadrupole moment of quark-gluon plasma in heavy ion collisions}",
    eprint = "1103.1307",
    archivePrefix = "arXiv",
    primaryClass = "hep-ph",
    doi = "10.1103/PhysRevLett.107.052303",
    journal = "Phys. Rev. Lett.",
    volume = "107",
    pages = "052303",
    year = "2011"
}

@article{Mao:2022dqn,
    author = "Mao, Shijun",
    title = "{Inverse catalysis effect of the quark anomalous magnetic moment to chiral restoration and deconfinement phase transitions at finite baryon chemical potential}",
    eprint = "2206.12054",
    archivePrefix = "arXiv",
    primaryClass = "nucl-th",
    doi = "10.1103/PhysRevD.106.034018",
    journal = "Phys. Rev. D",
    volume = "106",
    number = "3",
    pages = "034018",
    year = "2022"
}

@article{Bali:2011qj,
    author = "Bali, G. S. and Bruckmann, F. and Endrodi, G. and Fodor, Z. and Katz, S. D. and Krieg, S. and Schafer, A. and Szabo, K. K.",
    title = "{The QCD phase diagram for external magnetic fields}",
    eprint = "1111.4956",
    archivePrefix = "arXiv",
    primaryClass = "hep-lat",
    doi = "10.1007/JHEP02(2012)044",
    journal = "JHEP",
    volume = "02",
    pages = "044",
    year = "2012"
}

@article{Bali:2012zg,
    author = "Bali, G. S. and Bruckmann, F. and Endrodi, G. and Fodor, Z. and Katz, S. D. and Schafer, A.",
    title = "{QCD quark condensate in external magnetic fields}",
    eprint = "1206.4205",
    archivePrefix = "arXiv",
    primaryClass = "hep-lat",
    doi = "10.1103/PhysRevD.86.071502",
    journal = "Phys. Rev. D",
    volume = "86",
    pages = "071502",
    year = "2012"
}

@article{Bali:2014kia,
    author = {Bali, G. S. and Bruckmann, F. and Endr{\"o}di, G. and Katz, S. D. and Sch{\"a}fer, A.},
    title = "{The QCD equation of state in background magnetic fields}",
    eprint = "1406.0269",
    archivePrefix = "arXiv",
    primaryClass = "hep-lat",
    doi = "10.1007/JHEP08(2014)177",
    journal = "JHEP",
    volume = "08",
    pages = "177",
    year = "2014"
}

@article{Endrodi:2015oba,
    author = "Endrodi, Gergely",
    title = "{Critical point in the QCD phase diagram for extremely strong background magnetic fields}",
    eprint = "1504.08280",
    archivePrefix = "arXiv",
    primaryClass = "hep-lat",
    doi = "10.1007/JHEP07(2015)173",
    journal = "JHEP",
    volume = "07",
    pages = "173",
    year = "2015"
}

@article{Ferreira:2014kpa,
    author = "Ferreira, M. and Costa, P. and Louren{\c{c}}o, O. and Frederico, T. and Provid{\^e}ncia, C.",
    title = "{Inverse magnetic catalysis in the (2+1)-flavor Nambu-Jona-Lasinio and Polyakov-Nambu-Jona-Lasinio models}",
    eprint = "1404.5577",
    archivePrefix = "arXiv",
    primaryClass = "hep-ph",
    doi = "10.1103/PhysRevD.89.116011",
    journal = "Phys. Rev. D",
    volume = "89",
    number = "11",
    pages = "116011",
    year = "2014"
}

@article{Huang:2015oca,
    author = "Huang, Xu-Guang",
    title = "{Electromagnetic fields and anomalous transports in heavy-ion collisions --- A pedagogical review}",
    eprint = "1509.04073",
    archivePrefix = "arXiv",
    primaryClass = "nucl-th",
    doi = "10.1088/0034-4885/79/7/076302",
    journal = "Rept. Prog. Phys.",
    volume = "79",
    number = "7",
    pages = "076302",
    year = "2016"
}

@article{STAR:2009wot,
    author = "Abelev, B. I. and others",
    collaboration = "STAR",
    title = "{Azimuthal Charged-Particle Correlations and Possible Local Strong Parity Violation}",
    eprint = "0909.1739",
    archivePrefix = "arXiv",
    primaryClass = "nucl-ex",
    doi = "10.1103/PhysRevLett.103.251601",
    journal = "Phys. Rev. Lett.",
    volume = "103",
    pages = "251601",
    year = "2009"
}

@article{STAR:2009tro,
    author = "Abelev, B. I. and others",
    collaboration = "STAR",
    title = "{Observation of charge-dependent azimuthal correlations and possible local strong parity violation in heavy ion collisions}",
    eprint = "0909.1717",
    archivePrefix = "arXiv",
    primaryClass = "nucl-ex",
    doi = "10.1103/PhysRevC.81.054908",
    journal = "Phys. Rev. C",
    volume = "81",
    pages = "054908",
    year = "2010"
}

@article{ALICE:2012nhw,
    author = "Abelev, Betty and others",
    collaboration = "ALICE",
    title = "{Charge separation relative to the reaction plane in Pb-Pb collisions at $\sqrt{s_{NN}}= 2.76$ TeV}",
    eprint = "1207.0900",
    archivePrefix = "arXiv",
    primaryClass = "nucl-ex",
    reportNumber = "CERN-PH-EP-2012-183",
    doi = "10.1103/PhysRevLett.110.012301",
    journal = "Phys. Rev. Lett.",
    volume = "110",
    number = "1",
    pages = "012301",
    year = "2013"
}

@article{Wang:2021dcy,
    author = "Wang, Yuanyuan and Matsuzaki, Shinya",
    title = "{Axial inverse magnetic catalysis}",
    eprint = "2110.10432",
    archivePrefix = "arXiv",
    primaryClass = "hep-ph",
    doi = "10.1103/PhysRevD.105.074015",
    journal = "Phys. Rev. D",
    volume = "105",
    number = "7",
    pages = "074015",
    year = "2022"
}

@article{Ding:2026ewc,
    author = "Ding, Heng-Tong and Hern{\'a}ndez Hern{\'a}ndez, Jos{\'e} Javier and Zhang, Dan",
    title = "{Chiral and $U(1)_A$ symmetries in background magnetic fields from lattice QCD}",
    eprint = "2607.11625",
    archivePrefix = "arXiv",
    primaryClass = "hep-lat",
    month = "7",
    year = "2026"
}

@article{GomezNicola:2018pbx,
    author = "G{\'o}mez Nicola, A. and Ruiz De Elvira, J.",
    title = "{Chiral and $U(1)_A$ restoration for the scalar and pseudoscalar meson nonets}",
    eprint = "1803.08517",
    archivePrefix = "arXiv",
    primaryClass = "hep-ph",
    doi = "10.1103/PhysRevD.98.014020",
    journal = "Phys. Rev. D",
    volume = "98",
    number = "1",
    pages = "014020",
    year = "2018"
}

@article{Aoki:2012yj,
    author = "Aoki, Sinya and Fukaya, Hidenori and Taniguchi, Yusuke",
    title = "{Chiral symmetry restoration, eigenvalue density of Dirac operator and axial U(1) anomaly at finite temperature}",
    eprint = "1209.2061",
    archivePrefix = "arXiv",
    primaryClass = "hep-lat",
    reportNumber = "UTHEP-603, OU-HET-755",
    doi = "10.1103/PhysRevD.86.114512",
    journal = "Phys. Rev. D",
    volume = "86",
    pages = "114512",
    year = "2012"
}

@article{Buchoff:2013nra,
    author = "Buchoff, Michael I. and others",
    title = "{QCD chiral transition, U(1)A symmetry and the dirac spectrum using domain wall fermions}",
    eprint = "1309.4149",
    archivePrefix = "arXiv",
    primaryClass = "hep-lat",
    doi = "10.1103/PhysRevD.89.054514",
    journal = "Phys. Rev. D",
    volume = "89",
    number = "5",
    pages = "054514",
    year = "2014"
}

@article{Tomiya:2016jwr,
    author = "Tomiya, A. and Cossu, G. and Aoki, S. and Fukaya, H. and Hashimoto, S. and Kaneko, T. and Noaki, J.",
    title = "{Evidence of effective axial U(1) symmetry restoration at high temperature QCD}",
    eprint = "1612.01908",
    archivePrefix = "arXiv",
    primaryClass = "hep-lat",
    doi = "10.1103/PhysRevD.96.034509",
    journal = "Phys. Rev. D",
    volume = "96",
    number = "3",
    pages = "034509",
    year = "2017",
    note = "[Addendum: Phys.Rev.D 96, 079902 (2017)]"
}

@article{Mazur:2018pjw,
    author = "Mazur, Lukas and Kaczmarek, Olaf and Laermann, Edwin and Sharma, Sayantan",
    title = "{The fate of axial U(1) in 2+1 flavor QCD towards the chiral limit}",
    eprint = "1811.08222",
    archivePrefix = "arXiv",
    primaryClass = "hep-lat",
    doi = "10.22323/1.334.0153",
    journal = "PoS",
    volume = "LATTICE2018",
    pages = "153",
    year = "2019"
}

@article{Ding:2021jtn,
    author = "Ding, Heng-Tong and Li, Sheng-Tai and Wang, Xiao-Dan and Zhang, Yu and Tomiya, Akio and Mukherjee, Swagato",
    title = "{Correlated Dirac Eigenvalues and Axial Anomaly in Chiral Symmetric QCD}",
    eprint = "2112.00465",
    archivePrefix = "arXiv",
    primaryClass = "hep-lat",
    doi = "10.22323/1.396.0619",
    journal = "PoS",
    volume = "LATTICE2021",
    pages = "619",
    year = "2022"
}

@article{STAR:2021mii,
    author = "Abdallah, Mohamed and others",
    collaboration = "STAR",
    title = "{Search for the chiral magnetic effect with isobar collisions at $\sqrt {s_{NN}}$=200 GeV by the STAR Collaboration at the BNL Relativistic Heavy Ion Collider}",
    eprint = "2109.00131",
    archivePrefix = "arXiv",
    primaryClass = "nucl-ex",
    doi = "10.1103/PhysRevC.105.014901",
    journal = "Phys. Rev. C",
    volume = "105",
    number = "1",
    pages = "014901",
    year = "2022"
}

@article{Shuryak:1978ij,
    author = "Shuryak, Edward V.",
    title = "{Quark-Gluon Plasma and Hadronic Production of Leptons, Photons and Psions}",
    reportNumber = "IYF-78-24",
    doi = "10.1016/0370-2693(78)90370-2",
    journal = "Phys. Lett. B",
    volume = "78",
    pages = "150",
    year = "1978"
}

@article{STAR:2005gfr,
    author = "Adams, John and others",
    collaboration = "STAR",
    title = "{Experimental and theoretical challenges in the search for the quark gluon plasma: The STAR Collaboration's critical assessment of the evidence from RHIC collisions}",
    eprint = "nucl-ex/0501009",
    archivePrefix = "arXiv",
    doi = "10.1016/j.nuclphysa.2005.03.085",
    journal = "Nucl. Phys. A",
    volume = "757",
    pages = "102--183",
    year = "2005"
}

@article{Shuryak:2014zxa,
    author = "Shuryak, Edward",
    title = "{Strongly coupled quark-gluon plasma in heavy ion collisions}",
    eprint = "1412.8393",
    archivePrefix = "arXiv",
    primaryClass = "hep-ph",
    doi = "10.1103/RevModPhys.89.035001",
    journal = "Rev. Mod. Phys.",
    volume = "89",
    pages = "035001",
    year = "2017"
}

@article{Bzdak:2019pkr,
    author = "Bzdak, Adam and Esumi, Shinichi and Koch, Volker and Liao, Jinfeng and Stephanov, Mikhail and Xu, Nu",
    title = "{Mapping the Phases of Quantum Chromodynamics with Beam Energy Scan}",
    eprint = "1906.00936",
    archivePrefix = "arXiv",
    primaryClass = "nucl-th",
    doi = "10.1016/j.physrep.2020.01.005",
    journal = "Phys. Rept.",
    volume = "853",
    pages = "1--87",
    year = "2020"
}

@article{Kharzeev:2010gd,
    author = "Kharzeev, Dmitri E. and Yee, Ho-Ung",
    title = "{Chiral Magnetic Wave}",
    eprint = "1012.6026",
    archivePrefix = "arXiv",
    primaryClass = "hep-th",
    reportNumber = "BNL-94527-2010-JA",
    doi = "10.1103/PhysRevD.83.085007",
    journal = "Phys. Rev. D",
    volume = "83",
    pages = "085007",
    year = "2011"
}

@article{Kobayashi:1970ji,
    author = "Kobayashi, M. and Maskawa, T.",
    title = "{Chiral symmetry and eta-x mixing}",
    doi = "10.1143/PTP.44.1422",
    journal = "Prog. Theor. Phys.",
    volume = "44",
    pages = "1422--1424",
    year = "1970"
}

@article{Kobayashi:1971qz,
    author = "Kobayashi, M. and Kondo, H. and Maskawa, T.",
    title = "{Symmetry breaking of the chiral u(3) x u(3) and the quark model}",
    doi = "10.1143/PTP.45.1955",
    journal = "Prog. Theor. Phys.",
    volume = "45",
    pages = "1955--1959",
    year = "1971"
}

@article{tHooft:1976snw,
    author = "'t Hooft, Gerard",
    editor = "Shifman, Mikhail A.",
    title = "{Computation of the Quantum Effects Due to a Four-Dimensional Pseudoparticle}",
    reportNumber = "PRINT-76-0551 (HARVARD)",
    doi = "10.1103/PhysRevD.14.3432",
    journal = "Phys. Rev. D",
    volume = "14",
    pages = "3432--3450",
    year = "1976",
    note = "[Erratum: Phys.Rev.D 18, 2199 (1978)]"
}

@article{Fukushima:2001hr,
    author = "Fukushima, K. and Ohnishi, K. and Ohta, K.",
    title = "{Topological susceptibility at zero and finite temperature in the Nambu-Jona-Lasinio model}",
    eprint = "nucl-th/0101062",
    archivePrefix = "arXiv",
    doi = "10.1103/PhysRevC.63.045203",
    journal = "Phys. Rev. C",
    volume = "63",
    pages = "045203",
    year = "2001"
}

@article{Kunihiro:2009ds,
    author = "Kunihiro, Teiji",
    title = "{X Meson aka eta-prime and Kobayashi- Maskawa-'t Hooft Six-quark Vertex: U(1)(A) Anomaly and Generalized Nambu-Jona-Lasinio Model}",
    eprint = "0907.3808",
    archivePrefix = "arXiv",
    primaryClass = "hep-ph",
    reportNumber = "KUNS-2219",
    doi = "10.1143/PTP.122.255",
    journal = "Prog. Theor. Phys.",
    volume = "122",
    pages = "255--271",
    year = "2009"
}

@article{Fukushima:2003fw,
    author = "Fukushima, Kenji",
    title = "{Chiral effective model with the Polyakov loop}",
    eprint = "hep-ph/0310121",
    archivePrefix = "arXiv",
    reportNumber = "MIT-CTP-3424",
    doi = "10.1016/j.physletb.2004.04.027",
    journal = "Phys. Lett. B",
    volume = "591",
    pages = "277--284",
    year = "2004"
}

@article{Ratti:2005jh,
    author = "Ratti, Claudia and Thaler, Michael A. and Weise, Wolfram",
    title = "{Phases of QCD: Lattice thermodynamics and a field theoretical model}",
    eprint = "hep-ph/0506234",
    archivePrefix = "arXiv",
    doi = "10.1103/PhysRevD.73.014019",
    journal = "Phys. Rev. D",
    volume = "73",
    pages = "014019",
    year = "2006"
}

@article{Roessner:2006xn,
    author = "Roessner, Simon and Ratti, Claudia and Weise, W.",
    title = "{Polyakov loop, diquarks and the two-flavour phase diagram}",
    eprint = "hep-ph/0609281",
    archivePrefix = "arXiv",
    reportNumber = "ECT*-06-16",
    doi = "10.1103/PhysRevD.75.034007",
    journal = "Phys. Rev. D",
    volume = "75",
    pages = "034007",
    year = "2007"
}

@article{Bruckmann:2013oba,
    author = "Bruckmann, Falk and Endrodi, Gergely and Kovacs, Tamas G.",
    title = "{Inverse magnetic catalysis and the Polyakov loop}",
    eprint = "1303.3972",
    archivePrefix = "arXiv",
    primaryClass = "hep-lat",
    doi = "10.1007/JHEP04(2013)112",
    journal = "JHEP",
    volume = "04",
    pages = "112",
    year = "2013"
}

@article{Endrodi:2019zrl,
    author = "Endrodi, Gergely and Giordano, Matteo and Katz, Sandor D. and Kov{\'a}cs, T. G. and Pittler, Ferenc",
    title = "{Magnetic catalysis and inverse catalysis for heavy pions}",
    eprint = "1904.10296",
    archivePrefix = "arXiv",
    primaryClass = "hep-lat",
    doi = "10.1007/JHEP07(2019)007",
    journal = "JHEP",
    volume = "07",
    pages = "007",
    year = "2019"
}

@article{Ding:2022tqn,
    author = "Ding, H. -T. and Li, S. -T. and Liu, J. -H. and Wang, X. -D.",
    title = "{Chiral condensates and screening masses of neutral pseudoscalar mesons in thermomagnetic QCD medium}",
    eprint = "2201.02349",
    archivePrefix = "arXiv",
    primaryClass = "hep-lat",
    doi = "10.1103/PhysRevD.105.034514",
    journal = "Phys. Rev. D",
    volume = "105",
    number = "3",
    pages = "034514",
    year = "2022"
}

@article{Hattori:2023egw,
    author = "Hattori, Koichi and Itakura, Kazunori and Ozaki, Sho",
    title = "{Strong-field physics in QED and QCD: From fundamentals to applications}",
    eprint = "2305.03865",
    archivePrefix = "arXiv",
    primaryClass = "hep-ph",
    doi = "10.1016/j.ppnp.2023.104068",
    journal = "Prog. Part. Nucl. Phys.",
    volume = "133",
    pages = "104068",
    year = "2023"
}

@article{Endrodi:2024cqn,
    author = "Endrodi, Gergely",
    title = "{QCD with background electromagnetic fields on the lattice: A review}",
    eprint = "2406.19780",
    archivePrefix = "arXiv",
    primaryClass = "hep-lat",
    doi = "10.1016/j.ppnp.2024.104153",
    journal = "Prog. Part. Nucl. Phys.",
    volume = "141",
    pages = "104153",
    year = "2025"
}

@article{Fukushima:2012kc,
    author = "Fukushima, Kenji and Hidaka, Yoshimasa",
    title = "{Magnetic Catalysis Versus Magnetic Inhibition}",
    eprint = "1209.1319",
    archivePrefix = "arXiv",
    primaryClass = "hep-ph",
    doi = "10.1103/PhysRevLett.110.031601",
    journal = "Phys. Rev. Lett.",
    volume = "110",
    number = "3",
    pages = "031601",
    year = "2013"
}

@article{Mao:2016fha,
    author = "Mao, Shijun",
    title = "{Inverse magnetic catalysis in Nambu{\textendash}Jona-Lasinio model beyond mean field}",
    eprint = "1602.06503",
    archivePrefix = "arXiv",
    primaryClass = "hep-ph",
    doi = "10.1016/j.physletb.2016.05.018",
    journal = "Phys. Lett. B",
    volume = "758",
    pages = "195--199",
    year = "2016"
}

@article{Mao:2016lsr,
    author = "Mao, Shijun",
    title = "{From inverse to delayed magnetic catalysis in a strong magnetic field}",
    eprint = "1605.04526",
    archivePrefix = "arXiv",
    primaryClass = "hep-th",
    doi = "10.1103/PhysRevD.94.036007",
    journal = "Phys. Rev. D",
    volume = "94",
    number = "3",
    pages = "036007",
    year = "2016"
}

@article{Mao:2017tcf,
    author = "Mao, Shijun",
    title = "{Chiral Symmetry Restoration and Quark Deconfinement beyond Mean Field in a Magnetized PNJL Model}",
    eprint = "1712.06062",
    archivePrefix = "arXiv",
    primaryClass = "nucl-th",
    doi = "10.1103/PhysRevD.97.011501",
    journal = "Phys. Rev. D",
    volume = "97",
    number = "1",
    pages = "011501",
    year = "2018"
}

@article{Mao:2019avr,
    author = "Mao, Shijun",
    title = "{Chiral crossover characterized by Mott transition at finite temperature}",
    eprint = "1908.02851",
    archivePrefix = "arXiv",
    primaryClass = "nucl-th",
    doi = "10.1088/1674-1137/abcfad",
    journal = "Chin. Phys. C",
    volume = "45",
    number = "2",
    pages = "021004",
    year = "2021"
}

@article{Chao:2013qpa,
    author = "Chao, Jingyi and Chu, Pengcheng and Huang, Mei",
    title = "{Inverse magnetic catalysis induced by sphalerons}",
    eprint = "1305.1100",
    archivePrefix = "arXiv",
    primaryClass = "hep-ph",
    doi = "10.1103/PhysRevD.88.054009",
    journal = "Phys. Rev. D",
    volume = "88",
    pages = "054009",
    year = "2013"
}

@article{Farias:2014eca,
    author = "Farias, R. L. S. and Gomes, K. P. and Krein, G. I. and Pinto, M. B.",
    title = "{Importance of asymptotic freedom for the pseudocritical temperature in magnetized quark matter}",
    eprint = "1404.3931",
    archivePrefix = "arXiv",
    primaryClass = "hep-ph",
    doi = "10.1103/PhysRevC.90.025203",
    journal = "Phys. Rev. C",
    volume = "90",
    number = "2",
    pages = "025203",
    year = "2014"
}

@article{Gusynin:1995nb,
    author = "Gusynin, V. P. and Miransky, V. A. and Shovkovy, I. A.",
    title = "{Dimensional reduction and catalysis of dynamical symmetry breaking by a magnetic field}",
    eprint = "hep-ph/9509320",
    archivePrefix = "arXiv",
    reportNumber = "UCLA-95-TEP-26",
    doi = "10.1016/0550-3213(96)00021-1",
    journal = "Nucl. Phys. B",
    volume = "462",
    pages = "249--290",
    year = "1996"
}

@article{Ioffe:2006ww,
    author = "Ioffe, B. L.",
    title = "{Axial anomaly: The Modern status}",
    eprint = "hep-ph/0611026",
    archivePrefix = "arXiv",
    doi = "10.1142/S0217751X06035051",
    journal = "Int. J. Mod. Phys. A",
    volume = "21",
    pages = "6249--6266",
    year = "2006"
}

@article{Ding:2020xlj,
    author = "Ding, H. -T. and Li, S. -T. and Mukherjee, Swagato and Tomiya, A. and Wang, X. -D. and Zhang, Y.",
    title = "{Correlated Dirac Eigenvalues and Axial Anomaly in Chiral Symmetric QCD}",
    eprint = "2010.14836",
    archivePrefix = "arXiv",
    primaryClass = "hep-lat",
    doi = "10.1103/PhysRevLett.126.082001",
    journal = "Phys. Rev. Lett.",
    volume = "126",
    number = "8",
    pages = "082001",
    year = "2021"
}

@article{Costa:2004db,
    author = "Costa, Pedro and Ruivo, M. C. and de Sousa, C. A. and Kalinovsky, Yu. L.",
    title = "{Effective restoration of the U(A)(1) symmetry with temperature and density}",
    eprint = "hep-ph/0408177",
    archivePrefix = "arXiv",
    doi = "10.1103/PhysRevD.70.116013",
    journal = "Phys. Rev. D",
    volume = "70",
    pages = "116013",
    year = "2004"
}

@article{Ruivo:2011fg,
    author = "Ruivo, M. C. and Santos, M. and Costa, Pedro and de Sousa, C. A.",
    title = "{Interplay between chiral and axial symmetries in a SU(2) Nambu-Jona-Lasinio Model with the Polyakov loop}",
    eprint = "1112.6304",
    archivePrefix = "arXiv",
    primaryClass = "hep-ph",
    doi = "10.1103/PhysRevD.85.036001",
    journal = "Phys. Rev. D",
    volume = "85",
    pages = "036001",
    year = "2012"
}

@article{Jiang:2015xqz,
    author = "Jiang, Yin and Xia, Tao and Zhuang, Pengfei",
    title = "{Topological Susceptibility in Three-Flavor Quark Meson Model at Finite Temperature}",
    eprint = "1511.06466",
    archivePrefix = "arXiv",
    primaryClass = "hep-ph",
    doi = "10.1103/PhysRevD.93.074006",
    journal = "Phys. Rev. D",
    volume = "93",
    number = "7",
    pages = "074006",
    year = "2016"
}

@article{Li:2019chs,
    author = "Li, Xiang and Fu, Wei-Jie and Liu, Yu-Xin",
    title = "{New insight about the effective restoration of $U_A(1)$ symmetry}",
    eprint = "1910.05477",
    archivePrefix = "arXiv",
    primaryClass = "hep-ph",
    doi = "10.1103/PhysRevD.101.054034",
    journal = "Phys. Rev. D",
    volume = "101",
    number = "5",
    pages = "054034",
    year = "2020"
}

@article{Bjorken:1982qr,
    author = "Bjorken, J. D.",
    title = "{Highly Relativistic Nucleus-Nucleus Collisions: The Central Rapidity Region}",
    reportNumber = "FERMILAB-PUB-82-044-THY, FERMILAB-PUB-82-044-T",
    doi = "10.1103/PhysRevD.27.140",
    journal = "Phys. Rev. D",
    volume = "27",
    pages = "140--151",
    year = "1983"
}

@article{Gyulassy:2004zy,
    author = "Gyulassy, Miklos and McLerran, Larry",
    editor = "Rischke, D. and Levin, G.",
    title = "{New forms of QCD matter discovered at RHIC}",
    eprint = "nucl-th/0405013",
    archivePrefix = "arXiv",
    doi = "10.1016/j.nuclphysa.2004.10.034",
    journal = "Nucl. Phys. A",
    volume = "750",
    pages = "30--63",
    year = "2005"
}

@article{Voronyuk:2011jd,
    author = "Voronyuk, V. and Toneev, V. D. and Cassing, W. and Bratkovskaya, E. L. and Konchakovski, V. P. and Voloshin, S. A.",
    title = "{(Electro-)Magnetic field evolution in relativistic heavy-ion collisions}",
    eprint = "1103.4239",
    archivePrefix = "arXiv",
    primaryClass = "nucl-th",
    doi = "10.1103/PhysRevC.83.054911",
    journal = "Phys. Rev. C",
    volume = "83",
    pages = "054911",
    year = "2011"
}

@article{Son:2012wh,
    author = "Son, Dam Thanh and Yamamoto, Naoki",
    title = "{Berry Curvature, Triangle Anomalies, and the Chiral Magnetic Effect in Fermi Liquids}",
    eprint = "1203.2697",
    archivePrefix = "arXiv",
    primaryClass = "cond-mat.mes-hall",
    reportNumber = "INT-PUB-11-010",
    doi = "10.1103/PhysRevLett.109.181602",
    journal = "Phys. Rev. Lett.",
    volume = "109",
    pages = "181602",
    year = "2012"
}

@article{Zhao:2019hta,
    author = "Zhao, Jie and Wang, Fuqiang",
    title = "{Experimental searches for the chiral magnetic effect in heavy-ion collisions}",
    eprint = "1906.11413",
    archivePrefix = "arXiv",
    primaryClass = "nucl-ex",
    doi = "10.1016/j.ppnp.2019.05.001",
    journal = "Prog. Part. Nucl. Phys.",
    volume = "107",
    pages = "200--236",
    year = "2019"
}

@article{Landsteiner:2016led,
    author = "Landsteiner, Karl",
    title = "{Notes on Anomaly Induced Transport}",
    eprint = "1610.04413",
    archivePrefix = "arXiv",
    primaryClass = "hep-th",
    reportNumber = "IFT-UAM-CSIC-16-103",
    doi = "10.5506/APhysPolB.47.2617",
    journal = "Acta Phys. Polon. B",
    volume = "47",
    pages = "2617",
    year = "2016"
}

@article{Adler:1969gk,
    author = "Adler, Stephen L.",
    title = "{Axial vector vertex in spinor electrodynamics}",
    doi = "10.1103/PhysRev.177.2426",
    journal = "Phys. Rev.",
    volume = "177",
    pages = "2426--2438",
    year = "1969"
}

@article{Bell:1969ts,
    author = "Bell, J. S. and Jackiw, R.",
    title = "{A PCAC puzzle: $\pi^0 \to \gamma \gamma$ in the $\sigma$ model}",
    doi = "10.1007/BF02823296",
    journal = "Nuovo Cim. A",
    volume = "60",
    pages = "47--61",
    year = "1969"
}

@misc{Zhu2026Dataset,
  author       = {Zhu, X. and Sheng, X.-L. and Yang, S. and Mao, S. and Hou, D.},
  title        = {Dataset for axial symmetry breaking and chiral separation effect in NJL model},
  year         = {2026},
  publisher    = {Zenodo},
  doi          = {10.5281/zenodo.23062740},
  url          = {https://doi.org/10.5281/zenodo.23062740},
  note         = {[Dataset]}
}

\appendix
\renewcommand{\theequation}{\thesection.\arabic{equation}}
\renewcommand{\thefigure}{\thesection.\arabic{figure}}
\setcounter{equation}{0}
\setcounter{figure}{0}

\section{CSE conductivity in hadron gas and hybrid models}
\begin{figure}[th]
    \centering
    \includegraphics[width=0.9\linewidth]{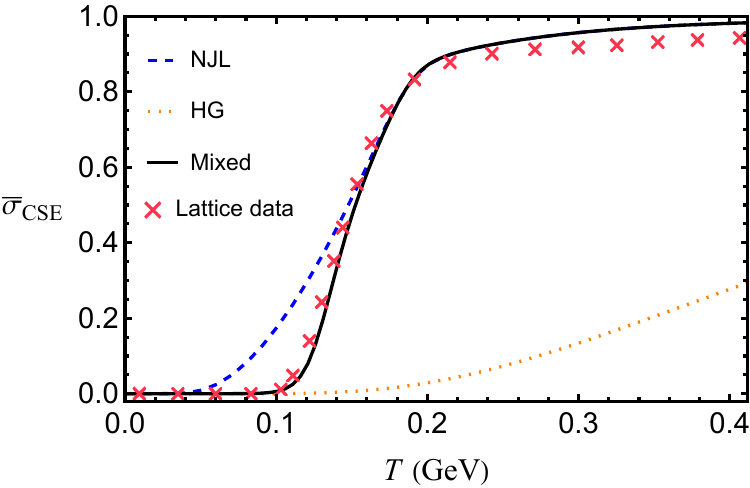}
    \caption{The normalized CSE conductivity $\bar{\sigma}_{\text{CSE}}$ as a function of $T$. The blue dashed and orange dotted lines show the results from the NJL and the hadron gas models, respectively, while the black solid line denotes the hybrid model. Lattice QCD data~\cite{Brandt:2023wgf} are shown for comparison. }
    \label{fig:CSE-mixed}
\end{figure}

The NJL model successfully reproduce the lattice-QCD results \citep{Brandt:2023wgf} at low or high temperatures but fail at intermediate temperature. The difference arises from the lack of quark confinement in the NJL. In order to reproduce the lattice-QCD results \citep{Brandt:2023wgf} across the entire temperature range, we construct a hybrid model mixing the NJL and the hadron gas (HG) models. The normalized CSE conductivity in the mixed model is constructed as follows
\begin{equation}\label{eq:mixed-model}
\left.\bar\sigma_\text{CSE}\right|_\text{Mixed}
=F(T)\left.\bar\sigma_\text{CSE}\right|_\text{NJL}
+\left[1-F(T)\right]\left.\bar\sigma_\text{CSE}\right|_\text{HG},
\end{equation}
where the weight function is set to the following form
\begin{equation}\label{eq:weight}
F(T)=\frac{1}{2}\bigg(1+\tanh\frac{T-T_{0}}{\sigma}\bigg)\,.
\end{equation}
Here $\left.\bar\sigma_\text{CSE}\right|_\text{NJL}$ is the result of the NJL model calculation discussed in Sec. \ref{sec:CSE} and $\left.\bar\sigma_\text{CSE}\right|_\text{HG}$ is the normalized CSE conductivity in the HG model given by
\begin{align} \label{eq:CSE-HG}
    \left.\bar\sigma_{\text{CSE}}(0,0,T)\right|_\text{HG}
    &=\left(\frac{2\pi^2}{\sum_H Q_H^2}\right)\sum_{H}\frac{Q_{H}^{2}}{4\pi^{2}T}
    \int dp_{z}\nonumber\\
    &\quad\times\left[\cosh\beta E_{H}^{(0)}+1\right]^{-1},
\end{align}
where $E_{H}^{(0)}=\sqrt{M_{H}^{2}+p_{z}^{2}}$ is lowest Landau level energy for the hadron flavor $H$ and 
$Q_H$ denotes its charge in units of $e$. Equation (\ref{eq:CSE-HG}) is normalized such that its chiral limit is 1. The sum runs over spin-1/2 hadrons including the proton, $\Sigma^+(1189)$, and $\Sigma^-(1197)$, while heavier baryons are neglected because their contributions are small. 
The parameters in the weight function (\ref{eq:weight}) are $T_{0}=0.13$ GeV and $\sigma=0.07$ GeV, which are determined by a least-squares fit to the lattice-QCD calculations \citep{Brandt:2023wgf}. As shown in Fig. \ref{fig:CSE-mixed}, this mixed model quantitatively agree with the lattice results in the whole temperature region.

\section{One-loop polarization matrices}\label{sec:one-loop}
In this section, we derive the one-loop polarization matrices in Eq. (\ref{eq:one-loop-polarization}). We start from a more general momentum-dependent definition
\begin{align}\label{eq:one-loop-Pi}
\Pi_{ff}^{S}(p) & \equiv -iN_{c}\int\frac{d^{4}q}{(2\pi)^{4}}\text{Tr}\left[G_{f}(q)G_{f}(q-p)\right]\,, \nonumber \\
\Pi_{ff}^{P}(p) & \equiv  -iN_{c}\int\frac{d^{4}q}{(2\pi)^{4}}\text{Tr}\left[i\gamma_{5}G_{f}(q)i\gamma_{5}G_{f}(q-p)\right]\,,
\end{align}
where $f=u,\,d,s\,$ denotes the flavor of quark. At finite chemical potential, the quark propagator under an external magnetic field along the $z$-direction reads \cite{Gusynin:1995nb} 
\begin{equation}\label{eq:quark-propagator}
G_{f}(q)=i e^{-q^2_{\perp}/|Q_{f}eB|}\sum_{n=0}^{\infty}
    \frac{(-1)^{n}D^{(n)}_{f}(q)}{(q_{0}+\mu_{f})^{2}-\left[E^{(n)}_{f}\right]^{2}+i\epsilon}\,,
\end{equation}
where $E^{(n)}_{f}\equiv\sqrt{M_f^2+q_z^2+2n|Q_feB|}$ is the on-shell quark energy.
The numerator is given by 
\begin{align}
D^{(n)}_{f}(q) =& 2\left[\gamma^{0}(q_{0}+\mu_{f})-\gamma^{3}q_{z}+M_{f}\right]\nonumber\\ &\times\Bigg[\mathcal{P}_{+}L^{(0)}_{n}\left(\frac{2q_{\perp}^{2}}{|Q_{f}eB|}\right)-\mathcal{P}_{-} L^{(0)}_{n-1}\left(\frac{2q_{\perp}^{2}}{|Q_{f}eB|}\right)\Bigg] 
    \nonumber\\
&-4(\gamma^1q_x+\gamma^2q_y)L^{(1)}_{n-1}\left(\frac{2q_{\perp}^{2}}{|Q_{f}eB|}\right)\,,
\end{align}
where $q_\perp=\sqrt{q_x^2+q_y^2}$ and the projection operators $\mathcal{P_{\pm}}=[1\pm i\text{sgn}(Q_{f}eB)\gamma^{1}\gamma^{2}]/2$.
Here $L^{(\alpha)}_n(x)$ is the generalized Laguerre polynomial with $L_{-1}^{(\alpha)}(x)\equiv0$.

Without loss of generality, we take $p^\mu=(p_0,{\bf 0})$. Substituting the quark propagator (\ref{eq:quark-propagator}) into $\Pi_{ff}^S(p)$ in (\ref{eq:one-loop-Pi}) we obtain 
\begin{align}
\Pi^{S}_{ff}(p)&=iN_{c}\int\frac{d^{4}q}{(2\pi)^{4}}e^{-2q_{\perp}^{2}/|Q_{f}eB|}\sum_{n,l}(-1)^{n+l}\nonumber\\
&\hspace{-1.2cm}\times\frac{\text{Tr}\left[D_{f}^{(n)}(q_{0}+\mu_f,{\bf q} )\,D_{f}^{(l)}(q_{0}+\mu_f-p_0,{\bf q})\right]}{\left[(q_{0}+\mu_f)^{2}-(E^{(n)}_f)^{2}\right]\left[(q_{0}+\mu_f-p_{0})^{2}-(E^{(l)}_f)^{2}\right]}\,.
\end{align}
Completing the trace of Dirac matrices and using the orthogonal relations of the generalized Laguerre polynomials,
\begin{equation}
\int_0^\infty dx\, x^\alpha e^{-x} L_n^{(\alpha)}(x)L_m^{(\alpha)}(x)=\frac{(n+\alpha)!}{n!}\delta_{mn}\,,
\end{equation}
we finally cast the result into the following form,
\begin{equation}
\Pi^S_{ff}(p)=2I_1(M_f)+\left[4M_f^2-(p_0)^2\right]I_2(M_f,p_0)\,.
\end{equation}
The auxiliary functions $I_1$ and $I_2$ are given by 
\begin{align}
I_{1}=&\frac{iN_{c}|q_{f}B|}{2\pi}\sum_{n}d_{(n)}\int\frac{dq_{0}dq_{z}}{(2\pi)^{2}}\frac{1}{(q_{0}+\mu_f)^2-(E^{(n)}_{f})^{2}}\nonumber\\
     =&-\frac{N_{c}|q_{f}B|}{2\pi}\sum_{n}d_{(n)}\int\frac{dp_{z}}{2\pi}\frac{1}{2E^{(n)}_{f}}\nonumber\\
     &\times\left[n^{+}_{f}(E^{(n)}_{f})+n^{-}_{f}(E^{(n)}_{f})-1\right]\,,
\end{align}
and 
\begin{align}
I_{2}=&\frac{iN_{c}|Q_{f}eB|}{2\pi}\sum_{n}d_{(n)}\int\frac{dq_{0}dq_{z}}{(2\pi)^{2}}\nonumber\\
&\times\left\{(q_{0}+\mu_f)^2-\left[E^{(n)}_{f}\right]^{2}\right\}^{-1}\nonumber\\
&\times\left\{(q_{0}+\mu_f-p_{0})^2-\left[E^{(n)}_{f}\right]^{2}\right\}^{-1}\,.
\end{align}
Completing the energy integral and taking the limit of $p_0\rightarrow0$ we obtain 
\begin{align}
    I_{2}=&\frac{N_{c}|q_{f}B|}{2\pi}\sum_{n}d_{(n)}\int\frac{dp_{z}}{2\pi}\nonumber\\
    &\times\Bigg\{\frac{n^{+}_{f}(E^{(n)}_{f})+n^{-}_{f}(E^{(n)}_{f})-1}{4(E^{(n)}_{f})^{3}}\nonumber\\
    &+\frac{n^{+}_{f}(E^{(n)}_{f})\left[1-n^{+}_{f}(E^{(n)}_{f})\right]}{4(E^{(n)}_{f})^{2}T}\nonumber\\
    &+\frac{n^{-}_{f}(E^{(n)}_{f})\left[1-n^{-}_{f}(E^{(n)}_{f})\right]}{4(E^{(n)}_{f})^{2}T}\Bigg\}\,.
\end{align}
Similarly, the polarization function of pseudo-scalar channel reads
\begin{equation}
    \Pi^{P}_{ff}(p)
     =2I_{1}(M_{f})-p_{0}^2
I_{2}(M_f,p_{0})\,.
\end{equation}
The one-loop polarization matrices in Eq. (\ref{eq:one-loop-polarization}) are then obtained by taking $p_0=0$ for $\Pi^S_{ff}(p)$ and $\Pi^P_{ff}(p)$. 

\end{document}